\documentclass[lettersize,journal]{IEEEtran}
\usepackage{dsfont}
\usepackage[font=small,center]{caption}
\usepackage{hyperref}
\usepackage{cite}
\usepackage{listings}
\usepackage{amsmath,amssymb,amsfonts}
\usepackage{amsthm}
\usepackage{algorithmic}
\usepackage{graphicx}
\usepackage{textcomp}
\usepackage{stmaryrd}
\usepackage{bbm}
\usepackage{bm}
\usepackage{amsmath}
\usepackage[utf8]{inputenc}
\usepackage[english]{babel}
\usepackage{mathtools} 
\usepackage{comment}
\usepackage{xcolor}

\usepackage{babel}

{}
\newtheorem{definition}{Definition}

\newtheorem*{remark}{Remark}
\usepackage{caption}
\usepackage{subcaption}
\usepackage{xcolor}
\usepackage{todonotes}

\definecolor{codegreen}{rgb}{0,0.6,0}
\definecolor{codegray}{rgb}{0.5,0.5,0.5}
\definecolor{codepurple}{rgb}{0.58,0,0.82}
\definecolor{backcolour}{rgb}{0.95,0.95,0.92}

\lstdefinestyle{mystyle}{
    backgroundcolor=\color{backcolour},   
    commentstyle=\color{codegreen},
    keywordstyle=\color{magenta},
    numberstyle=\tiny\color{codegray},
    stringstyle=\color{codepurple},
    basicstyle=\ttfamily\footnotesize,
    breakatwhitespace=false,         
    breaklines=true,                 
    captionpos=b,                    
    keepspaces=true,                 
    numbers=left,                    
    numbersep=5pt,                  
    showspaces=false,                
    showstringspaces=false,
    showtabs=false,                  
    tabsize=2
}

\DeclareMathOperator*{\argmax}{arg\,max}
\DeclareMathOperator*{\argmin}{arg\,min}
\allowdisplaybreaks
\usepackage{cite}
\usepackage{amsmath,amssymb,amsfonts}
\usepackage{algorithmic}
\usepackage{graphicx}
\usepackage{textcomp}
\usepackage{xcolor}
\def\BibTeX{{\rm B\kern-.05em{\sc i\kern-.025em b}\kern-.08em
    T\kern-.1667em\lower.7ex\hbox{E}\kern-.125emX}}
\usepackage{capt-of}
\usepackage{dsfont}
\usepackage{epsfig,latexsym,amsmath,amssymb,cite,color,graphicx,algorithm,algorithmic,cases}
\usepackage{amsmath,amssymb,amsfonts,hyperref,cite}

\usepackage{amsthm}
\usepackage{algorithmic}
\usepackage{graphicx}
\usepackage{setspace}
\usepackage[percent]{overpic}
\usepackage{stmaryrd}
\usepackage{bbm}
\usepackage{bm}
\usepackage{amsmath}
\usepackage[utf8]{inputenc}
\usepackage[english]{babel}
\usepackage{subcaption}

\hypersetup{
    colorlinks=true, 
    linkcolor= blue,
    citecolor=blue 
}    
    
\begin{document}

\title{Short Blocklength Wiretap Channel Codes via Deep Learning: Design and Performance Evaluation}
\author{\IEEEauthorblockN{Vidhi Rana and R\'emi A. Chou}\\
\thanks{V. Rana and R. Chou are with the Department of Electrical
Engineering and Computer Science, Wichita State University, Wichita, KS. E-mails: vxrana@shockers.wichita.edu, remi.chou@wichita.edu. This work was supported in part by NSF grant CCF-2047913. Part of this work has been presented at the 2021 IEEE Information Theory Workshop \cite{rana2021design}. 
}
}

\maketitle

\begin{abstract}
We design short blocklength codes for the Gaussian wiretap channel under information-theoretic security guarantees. Our approach consists in decoupling  the reliability and secrecy constraints in our code design. Specifically, we handle the reliability constraint via an autoencoder, and handle the secrecy constraint with hash functions. For blocklengths smaller than or equal to~$128$, we evaluate through simulations the probability of error at the legitimate receiver and the leakage at the eavesdropper for our code construction. This leakage is defined as the mutual information between the confidential message and the eavesdropper's channel observations, and is empirically measured via a neural network-based mutual information  estimator. 
Our simulation results provide examples of codes with positive secrecy rates that 
outperform the best known achievable secrecy rates obtained non-constructively for the Gaussian wiretap channel. 
Additionally, we show that our code design is suitable for the compound and arbitrarily varying Gaussian wiretap channels, for which the channel statistics are not perfectly known but only known to belong to a pre-specified uncertainty set. These models not only capture  uncertainty related to
channel statistics estimation, but also scenarios
where the eavesdropper jams the legitimate transmission or influences its own channel statistics by changing its~location.
\end{abstract}


\section{Introduction}
The wiretap channel \cite{wyner} is a basic model to account for eavesdroppers in  wireless communication. In this model, a sender (Alice) encodes a confidential message $M$ into
a codeword $X^n$ and transmits it to a legitimate
receiver (Bob) over $n$ uses of a channel in the presence of an  external eavesdropper (Eve).
 Bob's estimate of $M$ from his channel output observations is denoted by~$\hat{M}$, and Eve’s channel output observations are denoted by
$Z^n$. In \cite{wyner}, the constraints are that Bob must be able to recover $M$, i.e., $\lim_{n\rightarrow \infty}\mathbb{P}[M\neq \hat{M}]=0$, and the leakage about $M$ at Eve, quantified by  $I(M;Z^n)$, is not too large in the sense that $\lim_{n\rightarrow \infty}\frac{1}{n}I(M;Z^n)=0$. Note that the stronger security requirement $\lim_{n\rightarrow \infty}I(M;Z^n)=0$
 can also be considered~\cite{maurer}, meaning that Eve's observations~$Z^n$ are almost independent of $M$ for large $n$.
 The secrecy capacity has been characterized for degraded discrete memoryless channels in \cite{wyner}, for arbitrary  discrete memoryless channels in \cite{csiszar}, and for Gaussian channels in \cite{cheong}.

While \cite{wyner,csiszar,cheong} provide non-constructive achievability schemes for the wiretap channel, constructive coding schemes have also been proposed. Specifically, coding schemes based on  low-density parity-check (LDPC) codes \cite{thraj,subr,rathi}, polar codes~\cite{mahda,sasoglu,andersson,andersson2010nested}, and invertible extractors~\cite{hayashi,bellare} have been constructed for degraded or symmetric wiretap channel models. Moreover, the method in~\cite{bellare,hayashi} has been extended to the Gaussian wiretap channel~\cite{tyagi}. Coding schemes based on random lattice codes have also been proposed for the Gaussian wiretap channel \cite{ling}.
Subsequently, constructive \cite{wei,polar1,renes} and random \cite{gulcu} polar coding schemes  have been proposed to achieve the secrecy capacity of non-degraded discrete wiretap channels. Coding schemes that combine polar codes and invertible extractors have also been proposed to avoid the need for a pre-shared secret under strong secrecy~\cite{Profchou2020,chou2018explicit}.  
All the references above consider the asymptotic regime, i.e., the regime where $n$ approaches infinity. {\color{black}  However, many practical applications require short packet lengths or low latency \cite{ji2018}. To fulfill this need, non-asymptotic and second-order asymptotics achievability and converse bounds on the secrecy capacity of discrete and Gaussian wiretap channels  have been established in \cite{yang,hayashi2013,tan2012}.} Note that \cite{yang,hayashi2013,tan2012} focus on deriving fundamental limits and not on code constructions. We will review the works that are most related to our study and focus on code constructions at finite blocklength for the wiretap channel in Section \ref{relatedworks}.

In this paper, we propose to design short blocklength codes (smaller than or equal to~$128$) for the Gaussian wiretap channel under information-theoretic security guarantees. Such an information-theoretic approach enables coding solutions robust against computationally unbounded adversaries, and are thus technology independent and, in particular, quantum proof. 
 Specifically, we quantify security in terms of the leakage $I(M;Z^n)$, i.e., the mutual information between the confidential message and the eavesdropper's channel observations. 
The main idea of our approach is to  decouple the reliability and secrecy constraints. Specifically, we use a deep learning approach based on a feed-forward neural network autoencoder~\cite{goodfellow2016deep} to handle the reliability constraint and cryptographic tools, namely, hash functions \cite{carter}, to handle the secrecy constraint.\footnote{{\color{black} Note that a coding strategy that separately handles the reliability and secrecy constraints with two separate coding layers is also used  for the discrete wiretap channel in~\cite{hayashi,bellare}, and for the Gaussian wiretap channel in~\cite{tyagi}. In these works, an asymptotic regime is considered, i.e., the blocklength $n$ tends to infinity. Further, in \cite{hayashi,bellare,tyagi}, the security layer relies on the random choice of a hash function in a family of universal hash functions, and therefore, the coding scheme is non-constructive. In this paper, we also consider a family of hash functions for the security layer but only select a specific function in this family. This choice is deterministic and part of the coding scheme design, thus making it constructive, as elaborated on in our simulation~results.}} Then, to evaluate the performance of our constructed code, we empirically estimate the leakage~$I(M;Z^n)$. Note that even for small values of~$n$ this estimation is challenging with standard techniques such as binning of the probability space \cite{darbellay}, $k$-nearest neighbor statistics~\cite{kraskov}, or maximum likelihood estimation \cite{suzuki}. 
{\color{black}
     Unlike~\cite{yang,hayashi2013,tan2012}, which analytically derive upper bounds on the leakage, we consider a practical approach to estimate the leakage via the mutual information neural estimator (MINE) from~\cite{r3a}, which is provably consistent and offers better performances than other known mutual information estimators in high dimension. We also compare the performances of our codes with the best-known achievability and converse bounds on optimal secrecy rates for the Gaussian wiretap channel~\cite{yang}.}

{\color{black} Our main contributions are as follows.

\begin{enumerate} 
     \item 
     We propose a framework based on neural networks that enables a flexible design of finite blocklength codes for the Gaussian wiretap channel. Additionally, as seen in our simulations, our code design provides  examples of wiretap codes that outperform the best known achievable secrecy rates from  \cite{yang} obtained non-constructively for the Gaussian wiretap channel.
     
     \item We demonstrate that our proposed framework is also able to handle compound \cite{liang,bje2013} and arbitrarily varying \cite{bje2_2013,molavianjazi2009arbitrary} settings,  when uncertainty holds on both the legitimate users' channel and the eavesdropper's~channel, as demonstrated by our simulations results in Section \ref{comp_arb_wtc}. These models are particularly useful to capture uncertainty about the channel statistics of the eavesdropper channel or to model an active eavesdropper    who can influence its channel statistics by changing its location.

    \item We propose  a coding scheme design able to precisely control the level of information  leakage at the eavesdropper  through the independent design of a reliability coding layer and a secrecy coding layer. By contrast, as elaborated on in Section~\ref{relatedworks},  deep learning approaches that  seek to  simultaneously design codes for reliability and secrecy do not seem to offer good control over the information leakage at the eavesdropper.

\end{enumerate}
Additionally, our proposed code design offers the following features.
\begin{itemize}
   
\item A modular approach that separates the code design into a secrecy layer and a reliability layer. The secrecy layer only deals with the secrecy constraint and only depends on the statistics of the eavesdropper's channel, whereas the reliability layer only deals with the reliability constraint and only depends on the statistics of the legitimate receiver's channel.  This approach allows a simplified code design, for instance, if only one of the two layers needs to be (re)designed.

\item A universal way of dealing with the secrecy constraint through the use of hash functions.
This is beneficial, for instance,  for compound \cite{liang,bje2013} and arbitrarily varying~\cite{bje2_2013,molavianjazi2009arbitrary}
settings, as our results show that it becomes sufficient to design our code with respect to the
best eavesdropper’s channel.

\item A method that can be applied to an arbitrary channel model as the conditional probability distribution that defines the channel is not needed and only input and output channel samples are needed to design the reliability and secrecy layers.

\end{itemize}
Note that it is difficult to analytically characterize optimal secrecy rates for the Gaussian wiretap channel in the finite blocklength regime. In this study, we adopt a practical approach based on deep learning to better understand this regime.}

The remainder of the paper is organized as follows. Section~\ref{relatedworks} reviews related works. Section~\ref{model} introduces the Gaussian wiretap channel model. Section~\ref{codingscheme} describes our proposed code design and our simulation results for the Gaussian wiretap channel model. Section~\ref{comp_arb_wtc} discusses the  compound and arbitrarily varying Gaussian wiretap channel models and presents our simulation results. Finally,  Section~\ref{CR} provides   concluding~remarks. 

\section{Related Works}\label{relatedworks}

As elaborated on in the introduction, several code constructions have already been proposed for Gaussian wiretap channel coding in the asymptotic blocklength regime. Another challenging task is code designs in the finite blocklength regime. Next, we review known finite-length code constructions based on coding theoretic tools and deep learning tools in Sections~\ref{secA} and~\ref{secB},~respectively.

\subsection{Works based on coding theory} \label{secA}
In the following, we distinguish the works that consider a non-information-theoretic secrecy metric from the works that consider an information-theoretic secrecy metric.

\subsubsection{Non-information-theoretic secrecy metric}

A non-information-theoretic security metric called security gap, which is based on an error probability analysis at the eavesdropper, is used to evaluate the secrecy performance in~\cite{r2,noora,klinc,baldi,baldi2014secrecy}. Specifically, randomized convolutional codes for Gaussian and binary symmetric wiretap channels are studied in~\cite{r2}, and randomized turbo codes for the Gaussian wiretap channel are investigated in \cite{noora}. Coding schemes for the Gaussian wiretap channel based on LDPC codes are proposed in~\cite{klinc,baldi}. Additionally, another non-information-theoretic security approach called  practical secrecy is investigated in~\cite{harri}, where a leakage between Alice's message and an estimate of the message at Eve is estimated.

\subsubsection{Information-theoretic secrecy metric}
Next, we review works that consider the leakage~$I(M;Z^n)$ as  a secrecy metric. In~\cite{r4}, punctured systematic irregular LDPC codes are proposed for the binary phase-shift-keyed-constrained Gaussian wiretap channel, and 
a leakage as low as $11$ percent of the message length 
has been obtained for a blocklength~$n=10^6$.
In~\cite{baldi2}, LDPC codes   for the Gaussian wiretap channel have also been developed, and a leakage as low as $20$ percent of the message length has been obtained for a blocklength $n=50,000$. 
Most recently, in \cite{r1}, randomized Reed-Muller  codes are developed for the Gaussian wiretap channel, and a leakage as low as $0.2$ percent of the message length has been obtained for a blocklength $n=16$.

\subsection{Works based on deep learning}\label{secB}

Artificial neural networks have gained attention in communication system design because they approach the performance of state-of-the-art channel coding solutions. In \cite{oshea,dorner}, neural networks (autoencoder) are used  to learn  the encoder and decoder for a channel coding task without secrecy constraints. Other machine learning approaches for channel coding without secrecy constraints have also been investigated in \cite{aoudia, goutay} with reinforcement learning, in~\cite{r3} with mutual information estimators, and in~\cite{ye2018channel} with generative adversarial~networks. 

Recently, deep learning approaches for channel coding have been extended to wiretap channel coding. In \cite{gtc_frit,dl_frit}, a coding scheme that imitates coset coding by clustering learned signal constellations is developed for the Gaussian wiretap channel under a
non-information-theoretic secrecy metric, which relies on a cross-entropy loss function. {\color{black} In \cite{dl_mimo}, neural networks are used to learn optimal precoding for the MIMO Gaussian
wiretap channel.} 
In \cite{besser}, a coding scheme for the Gaussian wiretap channel is developed under the information-theoretic leakage $I(M;Z^n)$ with an autoencoder approach that seeks to  simultaneously optimize the reliability and secrecy constraints. A leakage as low as $15$ percent of the message length is obtained in \cite{besser}  for a blocklength $n=16$. It seems that precisely controlling and minimizing the leakage is challenging with such an approach. By contrast, in this paper, we propose an approach that separates the code design into a part that only deals with the reliability constraint (by means of an autoencoder) and another part that only deals with the secrecy constraint (by means of hash functions). As  supported by our simulation results, one of the advantages of our approach is a better control of how small the leakage can be made.

\section{Gaussian wiretap channel model}\label{model}
Notation: Unless specified otherwise, capital letters represent random variables,
whereas lowercase letters represent realizations of associated random variables, e.g., $x$ is a realization of the random variable~$X$.~$\vert \mathcal{X} \vert$  denotes the cardinality of the set $\mathcal{X}$. $\Vert \cdot \Vert_2$ denotes the Euclidean norm. $\textup{GF}(2^q )$ denotes a finite field of order $2^q$, $q \in \mathbb{N^*}$.
\smallskip

For $\mathcal{X}=\mathcal{Y}=\mathcal{Z}=\mathbb{R}$, consider a memoryless Gaussian  wiretap channel $(\mathcal{X},P_{YZ\vert X}, \mathcal{Y}\times\mathcal{Z})$ defined~by  
\begin{align}
    Y\triangleq X+N_{Y},\label{feb21_5_1} \\
    Z\triangleq X+N_{Z},\label{feb21_5_2}
\end{align}
where $N_{Y}$ and $N_{Z}$ are zero-mean Gaussian random variables with variances $\sigma^2_{Y}$ and $\sigma^2_{Z}$, respectively.
As formalized next, the objective of the sender is to transmit a confidential message~$M$ to a legitimate receiver by encoding it into a sequence~$X^n$, which is then sent over $n$ uses of the channels~\eqref{feb21_5_1}, \eqref{feb21_5_2} and yields the channel observations $Y^n$ and $Z^n$ at the legitimate receiver and eavesdropper, respectively.

\begin{definition} \label{def1}
Let $\mathbb{B}^n_0(\sqrt{nP})$ be the ball of radius $\sqrt{nP}$ centered at the origin in $\mathbb{R}^n$ under the Euclidian norm. An $(n,k,P)$ code consists of 
\begin{itemize}
    \item a message set $\{0,1\}^k$;
    \item an encoder $e:\{0,1\}^k\xrightarrow{} \mathbb{B}^n_0(\sqrt{nP})$, which, for a message $M \in \{0,1\}^k$, forms the codeword $X^n \triangleq e(M)$;
    \item a decoder $d: \mathbb{R}^n \xrightarrow{}\{0,1\}^k$, which, from the channel observations $Y^n$, forms an estimate of the message as $d(Y^n)$.
\end{itemize}
    The codomain of the encoder $e$ reflects the power constraint
$
    \Vert e(m)\Vert^2_2\leq nP,~\forall m\in \{0,1\}^k.
$
\end{definition}
The performance of an $(n,k,P)$ code is measured in terms~of 
\begin{enumerate}
       \item  The average probability of error
$    \mathbf{P_e} \triangleq\frac{1}{2^k}\sum_{m=1}^{2^k} \mathbb{P}[d(Y^n)\neq m\vert m~\text{is sent}]
$; \item The leakage at the eavesdropper
$
\mathbf{L_e}  \triangleq I(M;Z^n ).$
\end{enumerate}
\begin{definition}
An $(n,k,P)$ code is $\epsilon$-reliable if $\mathbf{P_e}\leq \epsilon$ and $\delta$-secure if $\mathbf{L_e}\leq \delta$. Moreover, a secrecy rate $\tfrac{k}{n}$ is  $(\epsilon,\delta)$-achievable with power constraint $P$ if there exists an $\epsilon$-reliable and $\delta$-secure $(n,k,P)$ code.
  \end{definition}

\section{Coding scheme}\label{codingscheme}

We first describe, at a high level, our coding scheme in Section \ref{cs_hld}. Specifically, our coding approach consists of two coding layers, one  reliability layer, whose design is described in {\color{black}Section}~\ref{rellyr}, and one security layer, whose design is described in Section \ref{SL}. {\color{black} We then comment on the communication rate of our proposed coding scheme when considering multiplexing of protected and unprotected messages in Section \ref{communR}.}
Finally, we provide simulation results and examples of our code design for the Gaussian wiretap channel in {\color{black}Sections~\ref{sim_siso}  and \ref{sim_lb}.} 

\subsection{High-level description of our coding scheme}\label{cs_hld}

Our code construction consists of (i) a reliability layer with an $\epsilon$-reliable $(n,q,P)$  code, described by the encoder/decoder pair $(e_0,d_0)$ (this code is designed without any security requirement, i.e., its performance is solely measured in terms of average probability of error),  and (ii) a security layer implemented with hash functions. We design the encoder/decoder pair $(e_0,d_0)$ of the reliability layer using a deep learning approach based on neural network autoencoders as described in Section \ref{rellyr}. We will then design two functions  $\varphi_s$ and $\psi_s$ in Section~\ref{SL} to perform the encoding and decoding, respectively, at the secrecy layer. The encoder/decoder pair~$(e,d)$ for the encoding and decoding process of the reliability and secrecy layers considered jointly is described as follows: 

{\it Encoding}: Assume that a fixed sequence of bits $s \in \mathcal{S} \triangleq \{0,1\}^q \backslash \{\mathbf{0}\}$, called seed, is known to all parties.   
Alice  generates a sequence $B$ of $q-k$ bits uniformly at random in $\{0,1\}^{q-k}$ (this sequence represents local randomness used to randomize the output of the function $\varphi_s$)  and encodes the message $M\in \{0,1\}^{k}$ as $e_0(\varphi_s(M,B))$,  where $\varphi_s(M,B) \in \{0,1\}^q$. The overall encoding map $e$ that describes the encoding at the secrecy and reliability  layers is described by  
\begin{align*}
    e:\{0,1\}^k\times \{0,1\}^{q-k}& \rightarrow{}\mathbb{B}^n_0(\sqrt{nP})\\ 
   (m,b) &\mapsto e_0(\varphi_s(m,b)).
\end{align*}

\begin{figure}[t]
\centering
\includegraphics[trim=5cm 3.2cm 5cm 3.5cm,clip,width=0.45\textwidth]{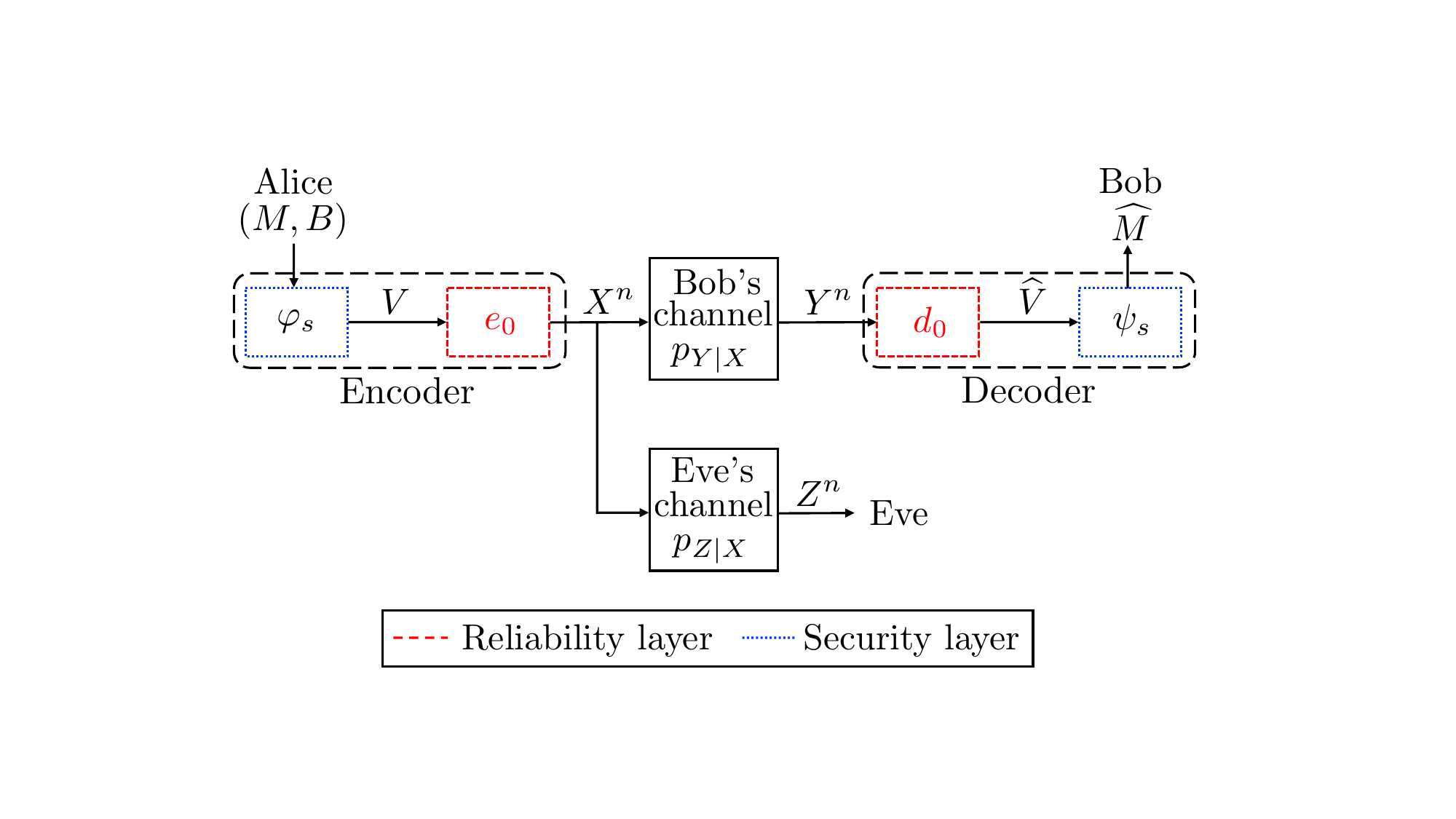}

\caption{Our code design  consists of a reliability layer and a security layer. The reliability layer is implemented using an autoencoder $(e_0,d_0)$ described in Section \ref{rellyr}, and the security layer is implemented using the functions $\varphi_s$ and $\psi_s$ described in Section \ref{SL1}.} \label{fig1}
\end{figure}

{\it Decoding}: Given $Y^n$ and $s$, Bob decodes the message as $\psi_s(d_0(Y^n))$. The overall decoding map $d$ that describes the decoding at the reliability and secrecy layers is 
\begin{align*}
   d:  \mathbb{R}^n &\rightarrow{}\{0,1\}^k \\
    y^n &\mapsto \psi_s(d_0(y^n)).
\end{align*}
For a given code design, described by the encoder/decoder pair $(e,d)$, we will then evaluate the performance of this code by empirically measuring the leakage using a neural network-based mutual information  estimator as described in Section~\ref{mine}. Our  code design is summarized in~Figure~\ref{fig1}.

\subsection{Design of the reliability layer $(e_0,d_0)$}\label{rellyr}

\begin{figure}[ht]
\centering
\includegraphics[trim=5.5cm 5.5cm 5cm 3.5cm,clip,width=0.5\textwidth]{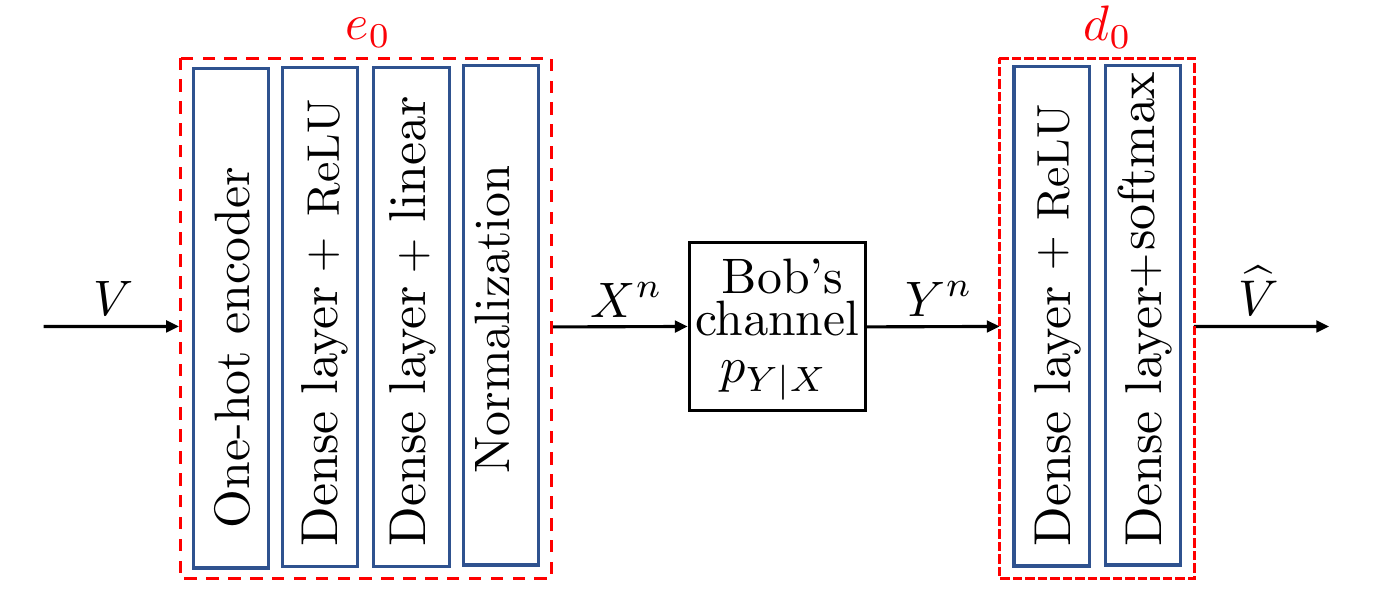}

\caption{Architecture of the autoencoder $(e_0,d_0)$ via feed-forward neural networks.} \label{autoencoder_fi}
\end{figure}
 
The design of the reliability layer consists in designing an $\epsilon$-reliable $(n,q,P)$ code described by the encoder/decoder pair $(e_0,d_0)$ for the channel \eqref{feb21_5_1}. Define $Q\triangleq 2^q$ and let $\mathcal{V}\triangleq\{1,2,\dots,Q\}$ be the message set of this code. { \color{black}  $(e_0,d_0)$ is obtained with an autoencoder as in \cite{oshea}. Specifically, the goal of the autoencoder is here
 to learn a representation of the encoded message that is robust to the channel noise, so that the received message at Bob can be reconstructed from its noisy channel observations with a small probability of error.  In other words,
the encoding part (denoted by $e_0$) of this autoencoder  adds redundancy to the message to ensure recoverability by Bob in the presence of noise.} As depicted in Figure~\ref{autoencoder_fi}, the  encoder consists of (i) {\color{black} a one-hot encoder} where the input $v \in \mathcal{V}$ is mapped to a one-hot vector~$\mathbf{1}_v\in\mathbb{R}^{Q}$, i.e., a vector whose components are all equal to zero except the $v$-th component which is equal to one, followed by (ii) dense hidden layers (with rectified linear unit (ReLU) or linear activation functions \cite{oshea}) that take $v$ as input and return an~$n$-dimensional vector, followed by (iii)~a normalization layer that ensures that the average power constraint $\frac{1}{n}\Vert  e_0(v)\Vert^2_2\leq P$ is met for the  codeword ${e}_0(v)$. Note that, without loss of generality, one can assume that $P=1$ since one can rewrite $\frac{1}{n}\Vert  e_0(v)\Vert^2_2\leq P$ as $\frac{1}{n}\Vert \tilde {e}_0(v)\Vert^2_2\leq 1$, where~$\tilde {e}_0(v)\triangleq e_0(v)/\sqrt{P}$. As depicted in Figure~\ref{autoencoder_fi}, the decoder consists of dense hidden layers and a softmax layer.  More specifically, let $\mu^{\vert\mathcal{V}\vert}$ be the output of the last dense layer in the decoder. The softmax layer takes $\mu^{\vert\mathcal{V}\vert}$ as input and returns a vector of probabilities $p^{\vert\mathcal{V}\vert}\in [0,1]^{\vert\mathcal{V}\vert}$,  whose components $p_v$, $v \in \mathcal{V}$, are
$
    p_v\triangleq \exp(\mu_v)\left(\sum_{i=1}^{\vert\mathcal{V}\vert}\exp(\mu_i)\right)^{-1}\!\!\!.
$ 
Finally, the decoded message  $\hat{v}$ corresponds to the index of the component of $p^{\vert\mathcal{V}\vert}$ associated with the  highest probability, i.e., $\hat{v}\in\argmax_{v\in \mathcal{V}}p_{v}$. 
The autoencoder is trained over all possible messages $v \in \mathcal{V}$ using a stochastic gradient descent (SGD)  \cite{adam} and the categorical cross-entropy loss function.

\subsection{Design of the security layer $(\varphi_s,\psi_s)$}\label{SL}  

 The objective is now to design $(\varphi_s,\psi_s)$ such that the total amount of leaked information about the original message is small in the sense that $I(M;Z^n)\leq \delta$, for some $\delta>0$. For a given choice of $(\varphi_s,\psi_s)$, the performance of our code construction will be evaluated using a mutual information neural estimator (MINE) \cite{r3a}. Before we describe the construction of $(\varphi_s,\psi_s)$, we review the definition of 2-universal hash functions. 
 
 \begin{definition}\cite{carter} 
Given two finite sets $\mathcal{X}$ and $\mathcal{Y}$, a family~$\mathcal{G}$ of functions from $\mathcal{X}$ to $\mathcal{Y}$ is~2-universal if
$ \forall x_1,x_2\in \mathcal{X},~x_1\neq x_2 \implies \mathbb{P}[G(x_1)=G(x_2)]\leq  \vert \mathcal{Y} \vert^{-1},
$
 where $G$ is the random variable that represents the choice of a function $g\in \mathcal{G}$ uniformly at random in $\mathcal{G}$.
\end{definition}
\subsubsection{Design of $(\varphi_s,\psi_s)$} \label{SL1}
Let $\mathcal{S}\triangleq\{0,1\}^q \backslash \{\mathbf{0}\}$. For $k\leq q$, consider the  2-universal hash family of functions $\mathcal{G}\triangleq\{\psi_s\}_{s\in \mathcal{S}}$, where for $s\in \mathcal{S}$,  
\begin{align*}
   \psi_s:\{0,1\}^q &\rightarrow\{0,1\}^k \\
    v &\mapsto(s\odot v)_k,
\end{align*}
where $\odot$ is the multiplication in $\textup{GF}(2^q)$ and $(\cdot)_k$ selects the $k$ most significant bits.  In our proposed code design, the security layer is handled via a specific function $\psi_s\in\mathcal{G}$ indexed by the seed $s\in \mathcal{S}$.  Then, we define
\begin{align}
 \varphi_s: \{0,1\}^k\times \{0,1\}^{q-k} &\rightarrow \{0,1\}^q \nonumber\\
    (m,b)& \mapsto s^{-1}\odot(m \Vert b),\label{eq_5622_1}
\end{align}
where $(\cdot\lVert\cdot)$ denotes the concatenation of two strings. 

When the secrecy layer is combined with the reliability layer, our coding scheme can be summarized as follows. The   input of the encoder $e_0$ is obtained by computing $V\triangleq \varphi_s(M, B)$, where $M\in \{0,1\}^k$ is the confidential message, and $B\in \{0,1\}^{q-k}$ is a sequence of $q-k$ random bits generated uniformly at random. After computing $V$, the encoder $e_0$, trained as described in Section~\ref{rellyr}, generates the codeword $X^n\triangleq e_0(V)$, which is sent over the channel by Alice. Bob and Eve observe $Y^n$ and $Z^n$, respectively, as described by \eqref{feb21_5_1} and \eqref{feb21_5_2}. The decoder $d_0$, trained as described in Section \ref{rellyr}, decodes $Y^n$ as $\widehat{V} \triangleq d_0(Y^n)$. Then, the receiver performs the multiplication of   $\widehat{V}$ and $s$, which is followed by a selection of the $k$ most significant bits to create an estimate $\widehat{M}$ of $M$, i.e., $\widehat{M}\triangleq \psi_s(\widehat{V})$.

\subsubsection{Leakage evaluation via Mutual Information Neural Estimator (MINE)\cite{r3a}}
\label{mine}

Let $\{T_{\theta}: \{0,1\}^k \times \mathbb{R}^n \to \mathbb{R}\}_{\theta\in\Theta}$ be the set of functions  parameterized by a deep neural network with parameters $\theta\in\Theta$. Define 
\begin{align*}
  I_{\Theta}(p_{MZ^n}) \!\triangleq \sup_{\theta \in \Theta}\mathbb{E}_{p_{MZ^n}}T_{\theta}(M, Z^n) 
    -\log \mathbb{E}_{p_Mp_{Z^n}}e^{T_{\theta}(M, Z^n)}\!,
\end{align*}
where  $p_{MZ^n}$ is the joint probability distribution of~$(M,Z^n)$. By \cite{r3a}, $I_{\Theta}(p_{MZ^n})$ can approximate the mutual information~$I(M;Z^n)$ with
arbitrary accuracy. Note that because the true distribution $p_{MZ^n}$ is unknown, one cannot directly use~$I_{\Theta}(p_{MZ^n})$ to estimate $I(M;Z^n)$. However, by estimating the expectations in $I_{\Theta}(p_{MZ^n})$ with samples from $p_{M{Z^n}}$ and~$p_M$ and $p_{Z^n}$, one can rewrite $I_{\Theta}(p_{MZ^n})$ as  
\begin{align*}
   \widehat{I}(M;Z^n)& \triangleq \sup_{\theta \in \Theta}\frac{1}{l} \sum_{i=1}^{l}T_{\theta}(m(i),z^n(i))\\&\phantom{------}-\log \left[\frac{1}{l} \sum_{i=1}^{l}e^{T_{\theta}(\bar{m}(i),\bar{z}^n(i))} \right],
\end{align*}
where the term  $\frac{1}{l} \sum_{i=1}^{l}T_{\theta}(m(i),z^n(i))$   represents a sample mean using $l$ samples $(m(i),z^n(i))_{i\in\{1, \ldots,l\}}$ from $p_{MZ^n}$, and the term $\frac{1}{l} \sum_{i=1}^{l}e^{T_{\theta}(m(i),z^n(i))}$ represents a sample mean using $l$ samples $(\bar{m}(i),\bar{z}^n(i))_{i\in\{1, \ldots,l\}}$ from $p_{M}p_{Z^n}$.

\begin{figure}[ht]
\centering
\includegraphics[width=0.48\textwidth]{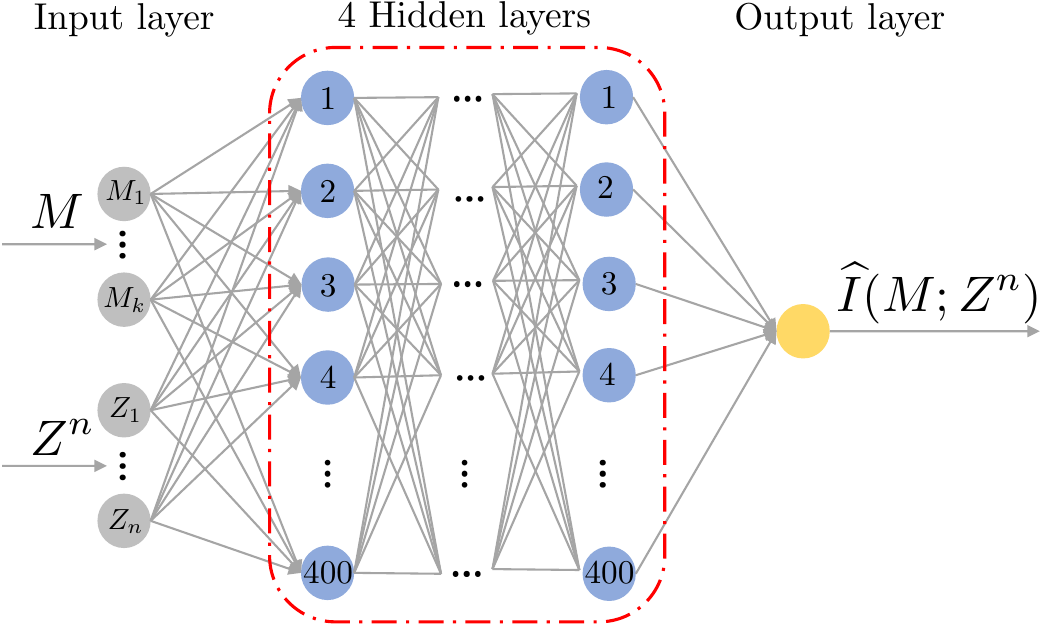}
\caption{The security performance is evaluated in terms of the leakage $I(M;Z^n)$ via the mutual information estimator described in Section \ref{mine}, where $M\triangleq (M_{i})_{i \in \{1,2,\dots,k\}}$, $Z^n\triangleq (Z_{j})_{j \in \{1,2,\dots,n\}}$.} \label{mine_fi}
\end{figure}

The goal of  MINE, whose architecture is depicted in Figure~\ref{mine_fi}, is to design $T_{\theta}$   such that $\widehat{I}(M;Z^n)$ approaches the mutual information  $I(M;Z^n)$. By \cite{r3a}, the estimator~$\widehat{I}(M;Z^n)$  converges to   $I(M;Z^n)$  when the number of samples is sufficiently large \cite{r3a}. Guidelines to implement the estimator~$\widehat{I}(M;Z^n)$  are also provided in \cite{r3a}.

\subsection{Discussion on the communication rate when multiplexing protected and unprotected messages}\label{communR}
       Note that our approach incurs no rate loss compared to a traditional channel code.
     Our proposed design of an $(n, k, P)$ code with power constraint $P$, blocklength $n$, and secret message length $k$ consists of two layers: (i) a reliability layer implemented with a $(n,q, P)$ channel code $(e_0,d_0)$, and (ii)  a secrecy layer. As described in Section~\ref{SL}, the secrecy layer takes as input a sequence of $q$ bits, out of which $k$ bits correspond to the secret message $M$ and $q-k$ bits correspond to random bits (denoted by $B$ in Section~\ref{SL}). By construction, the sequence $B$ can be reconstructed at Bob with an average probability of error $\mathbf{P}_e(e_0,d_0)$. However, the security constraint only holds on $M$ and not on $B$. To summarize, our code design transforms a channel code with rate $\frac{q}{n}$ into a wiretap code able to transmit a confidential message $M$ with rate $\frac{k}{n}$ and an unprotected message $B$ with rate $\frac{q-k}{n}$. Hence, there is no loss compared to a channel code, as the overall transmission rate is~$\frac{q}{n}$.
\subsection{Simulations and examples of code designs for $n\leq 16$}\label{sim_siso}

We now provide examples of code designs that follow the guidelines described in Sections~\ref{rellyr}, \ref{SL}, and evaluate their performance in terms of average probability of error at Bob and leakage at Eve.
The neural networks are implemented in Python 3.7 using Tensorflow~2.5.0.

\subsubsection{Autoencoder training for the design of the reliability layer $(e_0,d_0)$}\label{ae_tr}
We consider the channel
model (\ref{feb21_5_1}) with $\sigma^2_Y \triangleq 10^{-\text{SNR}_{B} / 10}$ and $\text{SNR}_{B}=9\textup{dB}$, where, as explained in Section~\ref{rellyr}, without loss of generality, we choose $P=1$. The autoencoder is trained for $q=n-1$ using SGD with the Adam optimizer~\cite{adam} at a learning
rate of $0.001$ over $600$ epochs of $10^5$ random encoder input messages with a batch size of $1000$. Due to the exponential growth of the complexity with~$q$, we changed the value of $q$ to $n-2$ when $n=16$. Specifically, to evaluate  $\mathbf{P}_e(e_0,d_0)$, we first generate the input $V\in\{0,1\}^q$. Then, $V$ is passed through the trained encoder $e_0$, which generates the codewords $X^n$ and the channel output $Y^n$. Finally, the trained decoder $d_0$ forms an estimate of~$V$ from $Y^n$.

\begin{figure}
\begin{center}
\begin{subfigure}[b]{0.55\textwidth}
\centering
   \begin{overpic}
[trim=3.5cm 8.5cm 2.5cm 9cm,clip,width=1\textwidth]{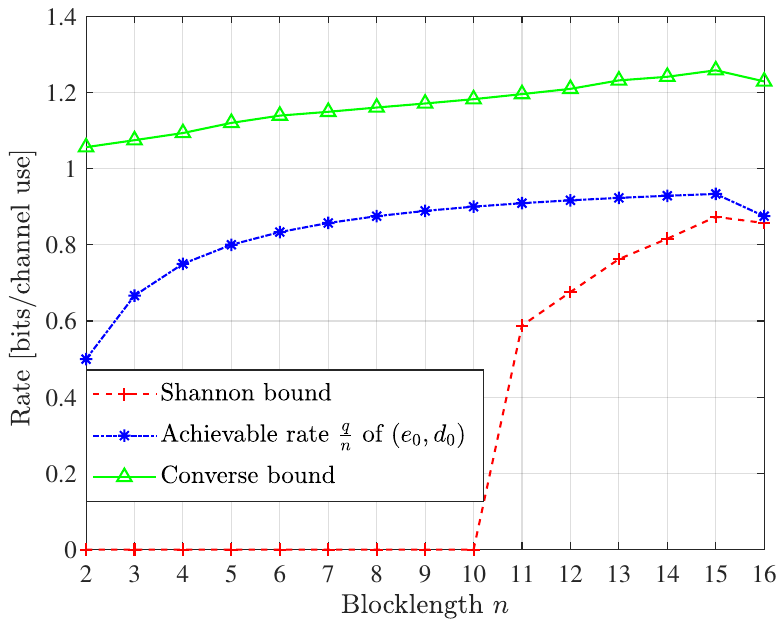}
    \put(39.3,23.6){\scriptsize [57, Eq.(41)]}
    \put(39.5,14.6){\scriptsize [57, Eq.(218)]}
 \end{overpic}
   \caption{}
   \label{figcomp_R} 
\end{subfigure}
\begin{subfigure}[b]{0.15\textwidth}
   \centering
   
 \includegraphics[trim=9.0cm 10.5cm 9cm 9.8cm,clip,width=1.0\textwidth]{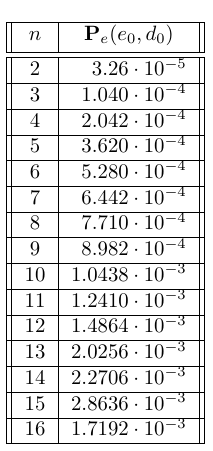}
   \caption{}
\label{tab:error_e0d0} 

\end{subfigure}

\caption[Comparisons]{Figure~\ref{figcomp_R} shows the rate versus the blocklength $n$ obtained with   $\epsilon\triangleq\mathbf{P}_e(e_0,d_0)$ listed in Figure~\ref{tab:error_e0d0} when $\text{SNR}_B=9\textup{dB}$.}

\end{center}
\end{figure}

{\color{black}Figure~\ref{figcomp_R} compares the achievable rate $\frac{q}{n}$ of our reliability layer $(e_0,d_0)$ with the best known achievability  and converse bounds \cite[Section III.J]{polyanskiy} for channel coding. We observe that the rate of our reliability layer outperforms the achievability bounds from \cite{polyanskiy} for blocklengths smaller than or equal to $16$ when $\text{SNR}_B=9\textup{dB}$. 
Note that for each value of $n$, this comparison is made for a given average probability of error~$\mathbf{P}_e(e_0,d_0)$ as specified in~Figure~\ref{tab:error_e0d0}.}

\subsubsection{Design of the secrecy layer and leakage evaluation}\label{sim_siso_lk}
\begin{figure}[ht]
\begin{center}
\includegraphics[trim=3cm 8cm 2.5cm 9cm,clip,width=0.55\textwidth]{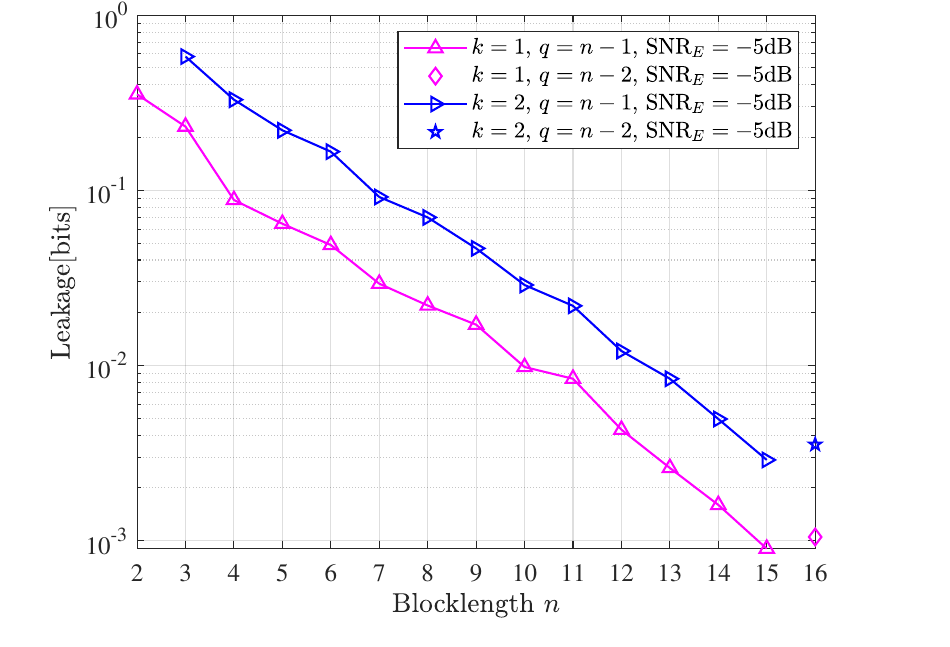}  
\caption{Leakage $\widehat{I}(M;Z^n)$ versus blocklength. When $n\in\{2,3,4\dots,15\}$, $q=n-1$, and when $n=16$, $q=n-2$.} \label{fig4}
\end{center}
\end{figure}

The seeds selected for the simulations are given in Table~\ref{tab:seeds}. {\color{black} All possible seeds have been tested for the values of $n$ smaller than or equal to eight to minimize the leakage, and only one seed is tested for the values of $n$ greater than eight. }
The leakage is evaluated using MINE as follows.
We used a fully connected feed-forward neural network with $4$ hidden layers, each having $400$ neurons, and a  ReLU as activation function. The input layer has $k+n$ neurons, and the Adam optimizer with a learning rate of $0.001$ is used for the training. The samples of the joint distribution $p_{MZ^n}$ are produced via uniform generation of messages $M\in \{0,1\}^k$ that are fed to the encoder $e =  e_0 \circ \varphi_s $, whose output  $X^n$ produces the channel output $Z^n$ at Eve. 
The samples of the marginal distributions are generated by dropping either $m$ or $z^n$ from the joint samples $(m,z^n)$. 
We have trained the neural network over $10000$ epochs of $20,000$ messages with a batch size of $2500$. 
Figure~\ref{fig4} shows the leakage versus the blocklength $n$ for different values of $k$ and $\text{SNR}_{E}= -5$dB.   {\color{black} We observe that the leakage increases as $k$ increases for fixed $n$ and $\text{SNR}_{E}$, which is also supported by the following inequality on the leakage. When $k=2$, if we write $M= (M_1,M_2)$, where $M_1, M_2 \in \{0,1\}$, then by the chain rule and nonnegativity of the mutual information, we have 
\begin{align*}
    I(M ;Z^n) =  
    I(M_1;Z^n)+I(M_2;Z^n\vert M_1)\nonumber \geq I(M_1;Z^n),
\end{align*}
where $I(M_1;Z^n)$ is interpreted as the leakage of a code with secrecy rate $\frac{1}{n}$ by considering that $M_2$ is a random bit  part of $B$ in (\ref{eq_5622_1}).
 
}
\begin{table}
\caption {Selected seeds for the security layer.} \label{tab:seeds} 
\begin{center}
\scalebox{0.8}{
 \begin{tabular}{||c | c |c||} 
 \hline
 $n$ &  seed $s$ & seed $s$\\&$(k=1)$ &$(k=2)$   \\ [0.5ex] 
 \hline\hline
 $2$ & 1  & - \\ 
 \hline
 $3$ & 11  & 11 \\
 \hline
 $4$ & 010 & 010 \\
 \hline
 $5$ & 1100 & 1100 \\
 \hline
 $6$ & 00010  & 00011 \\
 \hline
 $7$ & 001001  & 001001 \\
 \hline
 $8$ & 0001101 & 0001101 \\
 \hline
 $9$ & 10000000 & 10000000 \\
 \hline
 $10$ & 100000000 & 100000000 \\
 \hline
 $11$ & 1000000000 & 1000000000 \\
 \hline
 $12$ & 10000000000 & 10000000000 \\
 \hline
 $13$ & 100000000000 & 100000000000 \\
 \hline
 $14$ & 1000000000000 & 1000000000000 \\
 \hline
 $15$ & 10000000000000 & 10000000000000 \\
 \hline
 $16$ & 10000000000000 & 10000000000000 \\
 \hline
\end{tabular}
}
\end{center}
\end{table}

\begin{remark}
We observe a significant improvement in the leakage for a channel code coupled with our secrecy layer compared to the same channel code without any  secrecy layer. 
For instance, for the blocklength $n=8$ when $q=n-1$ and $\text{SNR}_E=-5\text{dB}$, the estimated mutual information between the input message of length $q$ to the encoder $e_0$ and the eavesdropper’s channel observations is  $\widehat{I}(V;Z^n)=1.55$ bits. 
Therefore, for a one-bit input, the leakage is 0.2214 bits on average. Also, for $n=8$, $q=n-1$, $k=1$,  $s=0001101$, and $\text{SNR}_E=-5\text{dB}$, the estimated mutual information between the one-bit confidential
message and the eavesdropper’s channel observations is $\widehat{I}(M;Z^n)=0.022$ bits. Hence, in this example, without our secrecy layer, a leakage as
low as $22$ percent is obtained per information bit on average, while with our secrecy layer, a leakage as
low as $2.2$ percent is obtained per information~bit.
 \end{remark}

\subsubsection{Average probability of error analysis}\label{av_rd}
 
\begin{figure}[h]
\begin{center}
\includegraphics[trim=3.5cm 8cm 2.5cm 9cm,clip,width=0.55\textwidth]{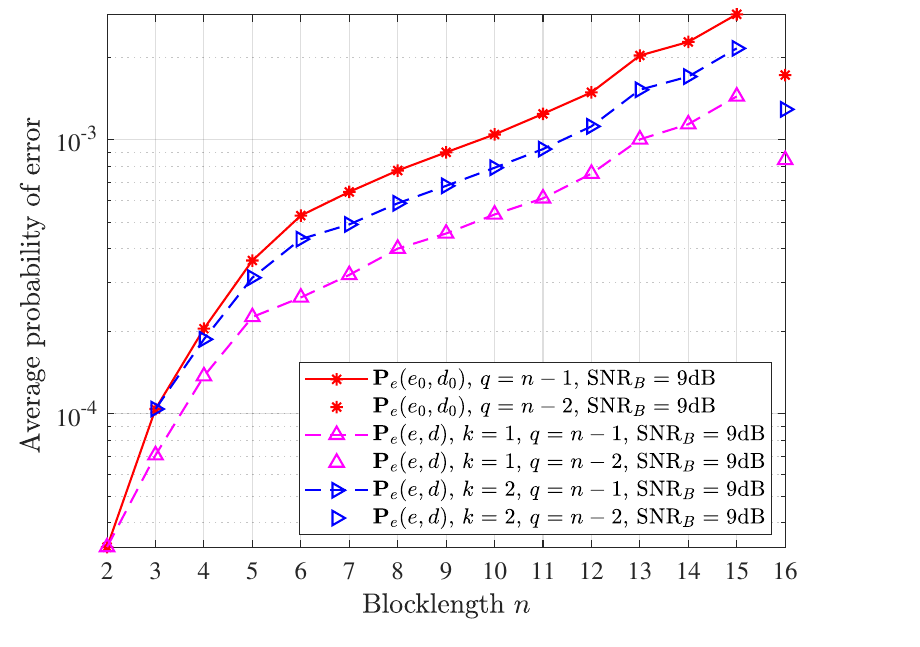} 
\caption{Average probability of error  versus blocklength. When $n\in\{2,3,4\dots,15\}$, $q=n-1$, and when $n=16$, $q=n-2$. During the training, the signal-to-noise ratio is $\text{SNR}_B=9$\textup{dB}. } \label{fig2}
\end{center}
\end{figure}

  To evaluate $\mathbf{P}_e(e,d)$, the trained encoder $e_0$ encodes the message $M \in \{0,1\}^k$ as $e_0(\varphi_s(M,B))$, as described in Section \ref{SL}, where $B\in \{0,1\}^{q-k}$ is a sequence of $q-k$ bits generated uniformly at random. The trained decoder $d_0$ forms $\widehat{M}\triangleq \psi_s(d_0(Y^n))$, as described in Section \ref{SL}. Figure~\ref{fig2} shows  $\mathbf{P}_e(e,d)$ versus the blocklength~$n$. Note that we only plotted $\mathbf{P}_e(e,d)$ when $k=1$ and $k=2$ as an example, as one will always have $\mathbf{P}_e(e,d) \leq \mathbf{P}_e(e_0,d_0)$ for any value~of~$k$ by construction. From Figure~\ref{fig2}, we also observe that, {for fixed $n, q$, and~$\mathbf{SNR}_B=9\textup{dB}$, the probability of error decreases as $k$ decreases, which is also supported by the following inequality 
     \begin{align*}
      \mathbb{P}[(\widehat{M}_1, \widehat{M}_2)\neq (M_1,M_2)]
      &\geq  \mathbb{P}[\widehat{M}_1\neq M_1],
  \end{align*}
where we write $M= (M_1,M_2)$ with $M_1, M_2 \in \{0,1\}$ when $k=2$ and $\mathbb{P}[\widehat{M}_1\neq M_1]$ is interpreted as the probability of error of a code with secrecy rate $\frac{1}{n}$ by considering that $M_2$ is a random bit  part of $B$ in~(\ref{eq_5622_1}).}

\begin{figure}
\begin{center}
\begin{subfigure}[b]{0.50\textwidth} 
\centering
    \begin{overpic}[trim=4.0cm 8.5cm 4.5cm 7cm,clip,width=0.90\textwidth]{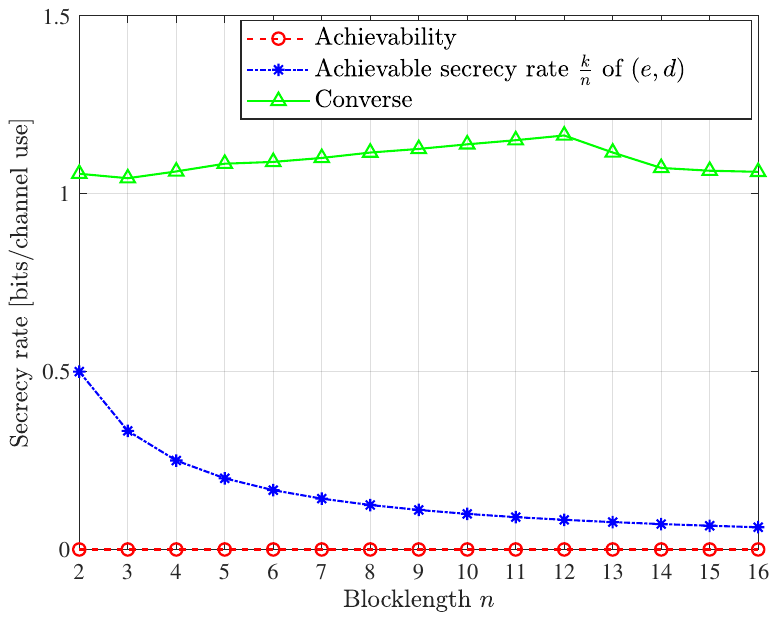}
   \put(53.8,66.1){\scriptsize [57, Eq. (218)], [24, Th. 12]}
   \put(60,73.8){\scriptsize  [24, Th. 7]}
 \end{overpic}
   \caption{}
   \label{figcomp_Rs} 
\end{subfigure}
\begin{subfigure}[b]{0.3\textwidth}
 \centering
 \includegraphics[trim=8.0cm 12.0cm 6.8cm 9.9cm,clip,width=0.83\textwidth]{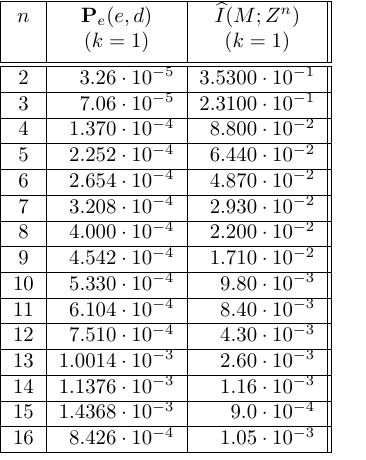} 
\caption {}  \label{tab:error_lk_ed} 
\end{subfigure}
\caption[Comparisons]{Figure~\ref{figcomp_Rs} shows the secrecy rate  versus the blocklength $n$  obtained from $\epsilon \triangleq \mathbf{P}_e(e,d)$ and $\delta\triangleq \widehat{I}(M;Z^n)$ listed in Figure~\ref{tab:error_lk_ed} when $\text{SNR}_B=9\textup{dB}$ and $\text{SNR}_E=-5\textup{dB}$. The converse bound is obtained as the minimum between \cite[Eq. (218)]{polyanskiy}  and \cite[Th. 12]{yang}.}
\end{center}
\end{figure}
\begin{figure}[h]
\begin{center}
\includegraphics[trim=3.5cm 8cm 2.5cm 8cm,clip,width=0.55\textwidth]{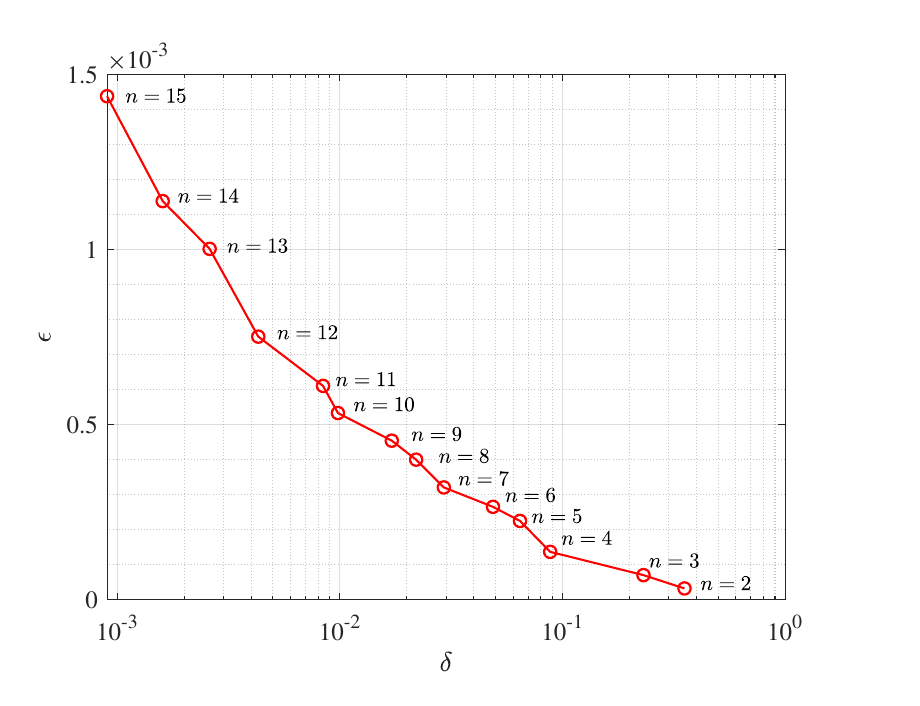} 
\caption{$\epsilon \triangleq \mathbf{P}_e(e,d)$ versus $\delta\triangleq \widehat{I}(M;Z^n)$ obtained from Figure~\ref{tab:error_lk_ed}.  When $n\in\{2,3,4\dots,15\}$, $q=n-1$, and the secrecy rate is~$\frac{1}{n}$. } \label{fig2_ach}
\end{center}
\end{figure}
\subsubsection{Discussion}
From Figures~\ref{fig4} and~\ref{fig2}, we, for instance, see that for $\text{SNR}_{B}=9$\textup{dB} and $\text{SNR}_{E}=-5$\textup{dB}, we have designed codes that  show that  the secrecy rate $\frac{1}{10}$ is~$(\epsilon= 5.330 \cdot 10^{-4},\delta=9.80\cdot10^{-3})$-achievable with blocklength $n=10$,  
and the secrecy rate $\frac{2}{13}$   is~$(\epsilon=1.5194\cdot 10^{-3},\delta= 8.40 \cdot 10^{-3})$-achievable with blocklength~$n=13$. 

{Figure~\ref{figcomp_Rs} compares the achievable secrecy rate $\frac{k}{n}$ of our code design $(e,d)$ with the best known achievability \cite[Theorem 7]{yang} and converse bounds \cite[Theorem 12]{yang} for the Gaussian wiretap channel, which are reviewed in the appendix for convenience. We observe that the rate of our code outperforms the best known achievability bounds for blocklengths smaller than or equal to $16$ when~$k=1$,~$\text{SNR}_B=9\textup{dB}$,~and~$\text{SNR}_E=-5\textup{dB}$. 
Note that the best known upper bounds from~\cite{yang} may not be optimal for small blocklengths, and improving them is an open problem. 
Note also that for each value of $n$, the comparison is made for a given average probability of error $\mathbf{P}_e(e,d)$ and leakage $\widehat{I}(M;Z^n)$ as specified in Figure~\ref{tab:error_lk_ed}. }

In Figure \ref{fig2_ach}, we also plotted  $\epsilon \triangleq \mathbf{P}_e(e,d)$ versus $\delta\triangleq \widehat{I}(M;Z^n)$ obtained from Figure~\ref{tab:error_lk_ed}.

\subsection{Simulations and examples of code designs for $n\leq 128$}\label{sim_lb}

\begin{figure}
\begin{center}
\includegraphics[trim=3.5cm 8cm 2.5cm 9cm,clip,width=0.55\textwidth]{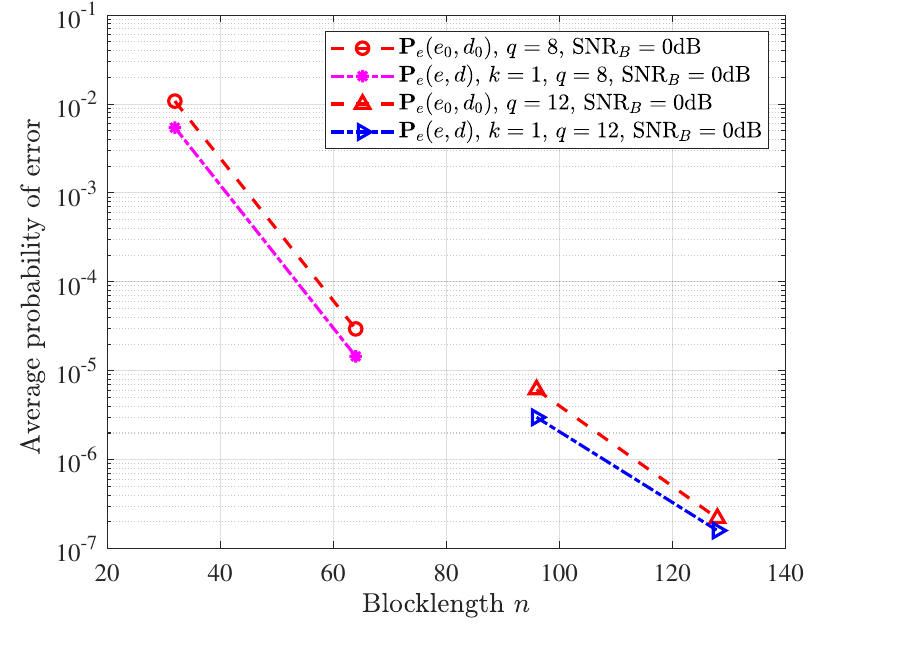}  
\caption{Average probability of error  versus blocklength. When $n\in\{32,64\}$, $q=8$, and when $n\in\{96, 128\}$, $q=12$.} \label{figerror_ae}
\end{center}
\end{figure}

We consider the channel model \eqref{feb21_5_1} with~$\sigma^2_Y~\triangleq~10^{-\text{SNR}_{B} / 10}$ and $\text{SNR}_{B}=0\textup{dB}$. For the design of the reliability layer $(e_0,d_0)$, the autoencoder is trained for $(n,q)=(32,8)$, $(n,q)=(64,8)$, $(n,q)=(96,12)$, and $(n,q)=(128,12)$  at a learning rate of $0.0001$ over $600$ epochs of $ 4\times 10^5$ random encoder input messages with a batch size of $5000$. Then, $50\times 10^6$ random messages are used to evaluate the average probability of errors $\mathbf{P}_e(e_0,d_0)$ and $\mathbf{P}_e(e,d)$ as described in Sections~\ref{ae_tr} and \ref{av_rd}. Figure \ref{figerror_ae} shows $\mathbf{P}_e(e,d)$ versus the blocklength $n$. As expected, we observe that, for fixed $k$ and $q$, the probability of error decreases as $n$ increases.

\begin{figure}
\begin{center}
\includegraphics[trim=3.5cm 8cm 2.5cm 9cm,clip,width=0.55\textwidth]{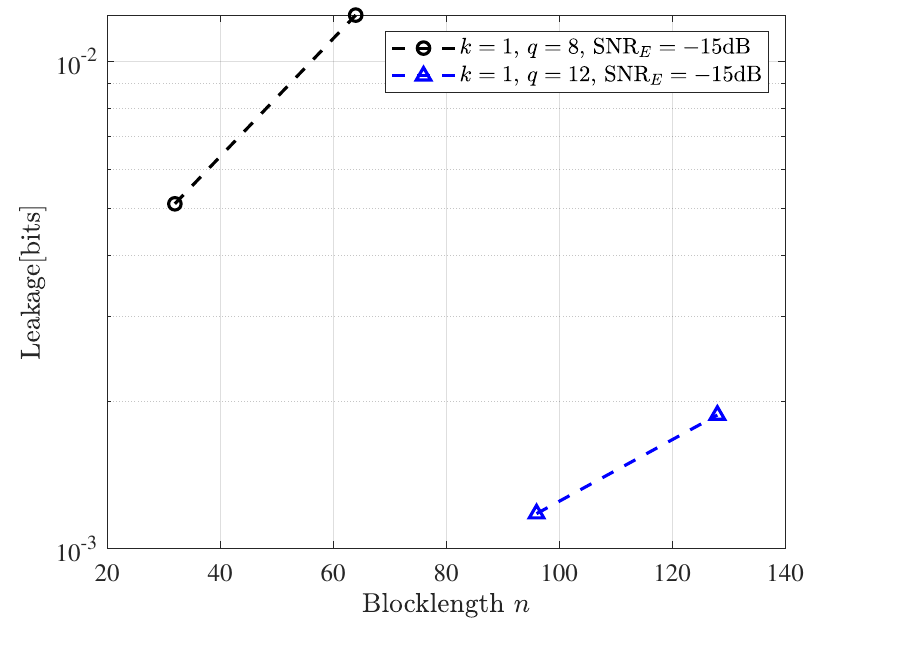}  
\caption{Leakage $\widehat{I}(M;Z^n)$ versus blocklength. When $n\in\{32,64\}$, $q=8$, and when $n\in\{96,128\}$, $q=12$.} \label{figlk_ae}
\end{center}
\end{figure}

The secrecy layer is implemented similarly to Section \ref{SL} with $k=1$, and we compute the leakage $I(M;Z^n)$ as in Section \ref{sim_siso_lk}.  We consider the model in \eqref{feb21_5_2} with~$\sigma^2_Z~\triangleq~10^{-\text{SNR}_{E} / 10}$ and $\text{SNR}_{E}=-15\textup{dB}$. Additionally, in our simulations, for blocklengths $n=32$ and $n=64$, the seed is chosen as $s=00000001$, and for blocklengths $n=96$ and $n=128$, the seed is chosen as $s=000010000000$. Figure~\ref{figlk_ae} shows  $\widehat{I}(M;Z^n)$ versus the blocklength $n$. 
As expected, we observe that, for fixed $k$ and $q$, the leakage increases as $n$ increases.
\begin{figure}
\begin{center}
\begin{subfigure}[b]{0.50\textwidth} 
\centering
   \begin{overpic}[trim=3cm 8cm 2.5cm 9cm,clip,width=1.1\textwidth]{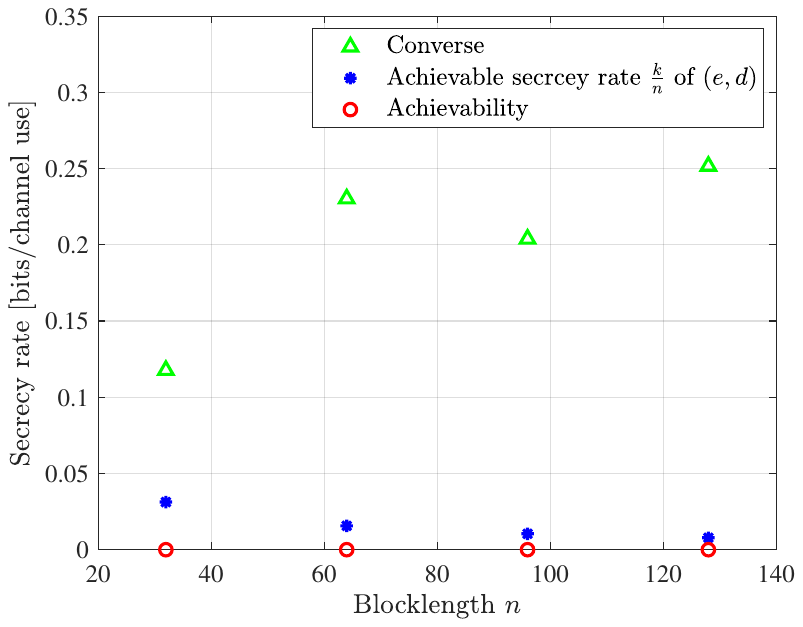}
   \put (56,62.5){\scriptsize [24, Eq. (116)]}
   \put (60,56){\scriptsize [24, Th. 7]}
   \end{overpic}
   \caption{}
   \label{lb_bounds}
\end{subfigure}
\begin{subfigure}[b]{0.30\textwidth}
   \centering
 \includegraphics[trim=7.0cm 14.3cm 6.5cm 12cm,clip,width=1\textwidth]{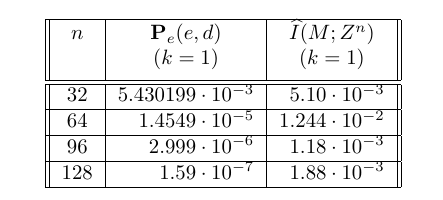} 
\caption {}  \label{tab:error_lk_ed_lb} 
\end{subfigure}
\caption[Comparisons]{Figure~\ref{lb_bounds} shows the secrecy rate  versus the blocklength $n$  obtained from $\epsilon \triangleq \mathbf{P}_e(e,d)$ and $\delta\triangleq \widehat{I}(M;Z^n)$ listed in Figure~\ref{tab:error_lk_ed_lb} when $\text{SNR}_B=0\textup{dB}$ and $\text{SNR}_E=-15\textup{dB}$. }
\end{center}
\end{figure}
Finally, we observe in Figure \ref{lb_bounds} that the rate of our code outperforms the best known achievability bounds~\cite[Theorem 7]{yang} for blocklengths smaller than or equal to $128$ when~$k=1$,~$\text{SNR}_B=0\textup{dB}$,~and~$\text{SNR}_E=-15\textup{dB}$.

\section{Compound and arbitrarily varying wiretap channel models}\label{comp_arb_wtc}
We first motivate the compound and arbitrarily varying wiretap channel models in Section~\ref{bg_comp_ar}. We then formally describe these two models in Section \ref{c_a_mo}. We present our coding scheme design in  Section \ref{c_a_de}. Finally, we evaluate the performances of our code design through simulations for 
 the compound and arbitrarily varying Gaussian wiretap channels in Sections~\ref{comp_s} and \ref{arb_s}, respectively. 
\subsection{Background} \label{bg_comp_ar}
In the setting of Section \ref{model}, the channel statistics are assumed to be perfectly known to Alice and Bob and fixed during the entire transmission. However, in practice, the channel statistics may not be perfectly known due to the nature of the wireless channel and inaccuracy in estimating channel statistics. Further, in some scenarios,  eavesdroppers could be active and influence their own channel statistics by changing their location, or the statistics of Bob's channel through jamming. To model such scenarios, two types of models have been introduced: compound wiretap channels and arbitrarily varying wiretap channels. For compound models, e.g.,~\cite{liang,bje2013,schaefer2015,boche2015}, the channel statistics are  fixed for all channel uses. Whereas for arbitrarily varying models, e.g.,~\cite{bje2_2013,schaefer2015,boche2015,chou2017gaussian,molavianjazi2009arbitrary,chou2022gaussian}, the channel statistics may change in an unknown and arbitrary manner from channel use to channel use.
Constructive coding schemes have also been proposed in~\cite{Profchou2020,chou2018explicit} for discrete compound and arbitrarily varying wiretap channels. While all the works above consider the asymptotic regime, in this section, we design short blocklength codes for the compound and arbitrarily varying wiretap channel models. 
\subsection{Models}\label{c_a_mo}
 
For $\mathcal{X}=\mathcal{Y}=\mathcal{Z}=\mathbb{R}$, a  compound or arbitrarily varying memoryless  Gaussian wiretap channel $\big(\mathcal{X},(p_{Y
_iZ_j\vert X})_{i\in \mathcal{I}, j\in \mathcal{J}}, \mathcal{Y}\times\mathcal{Z}\big)$ 
is defined for $i\in \mathcal{I}$, $j\in \mathcal{J}$,~by
\begin{align*}
    Y_i\triangleq X+N_{Y_i},\quad 
    Z_j\triangleq X+N_{Z_j},
\end{align*}
where $N_{Y_i}$ and $N_{Z_j}$ are zero mean Gaussian random variables with variances $\sigma^2_{Y_i}$ and $\sigma^2_{Z_j}$, respectively.

For the compound wiretap channel model, the channel statistics are constant throughout the transmission and are known to belong to given uncertainty sets $\mathcal{I}$, $\mathcal{J}$. The confidential message $M$ is encoded into a transmitted sequence $X^n$, and $Y^n_i$ and $Z^n_j$  represent the corresponding received sequence at the legitimate receiver and eavesdropper, respectively, for some $i \in \mathcal{I}$ and~$j \in \mathcal{J}$.  
\begin{definition}
A secrecy rate $\tfrac{k}{n}$ is $(\epsilon,\delta)$-achievable with power constraint $P$ for the compound wiretap channel if there exists a $(n,k,P)$ code such that
\begin{align}
    \max_{i \in \mathcal{I}}\mathbf{P}^i_e(e,d) &\leq \epsilon, \label{eqc1}\\
    \max_{j\in \mathcal{J}}I(M;Z_j^n ) &\leq \delta,  \label{eqc2}
\end{align}
where $\mathbf{P}^i_e(e,d)\triangleq\frac{1}{2^k}\sum_{m=1}^{2^k} \mathbb{P}[d(Y_{i}^n)\neq m\vert m~\text{is sent}]$.
  \end{definition}

In contrast to the compound wiretap channel, the channel statistics in arbitrarily varying wiretap channel models may vary in an unknown and arbitrary manner from channel use to channel use. Specifically, for the arbitrarily varying wiretap channel model, let $Y^n_{\mathbf{i}}$ and $Z^n_{\mathbf{j}}$  represent the corresponding received sequence at the legitimate receiver and eavesdropper, respectively, for some  $\mathbf{i} \in \mathcal{I}^n$ and $\mathbf{j} \in \mathcal{J}^n$.

\begin{definition}
A secrecy rate $\tfrac{k}{n}$ is   $(\epsilon,\delta)$-achievable with power constraint $P$ for the arbitrarily varying wiretap channel if there exists a $(n,k,P)$ code such that
\begin{align*}
     \max_{\mathbf{i} \in \mathcal{I}^n}\mathbf{P}^{\mathbf{i}}_e(e,d)  &\leq \epsilon,\\
   \max_{\mathbf{j}\in \mathcal{J}^n}I(M;Z_{\mathbf{j}}^n ) &\leq \delta,
\end{align*}
where $\mathbf{P}^{\mathbf{i}}_e(e,d)\triangleq\frac{1}{2^k}\sum_{m=1}^{2^k} \mathbb{P}[d(Y_{\mathbf{i}}^n)\neq m\vert m~\text{is sent}]$.
  \end{definition}

\subsection{Coding scheme design}\label{c_a_de}
For the compound and the arbitrarily varying wiretap channels, the design of $(e_0,d_0)$ for the reliability layer and $(\varphi_s,\psi_s)$ for the secrecy layer is similar to Sections \ref{rellyr} and \ref{SL}, respectively. Specifically, we train the encoder/decoder pair for Bob's channel with noise variance $\sigma^2_{Y_{i^*}}\triangleq\max_{i\in\mathcal{I}}\sigma^2_{Y_i}$, where $\sigma^2_{Y_i}\triangleq  10^{-\text{SNR}_B(i)/10}$, $i\in \mathcal{I}$. In other words, the reliability layer is designed for the worse, in terms of signal-to-noise ratio, Bob's channel. Note also that, during the training phase, the noise variance is fixed for all the channel uses. Then, we optimize the seed $s$ by minimizing the leakage for Eve's channel with  noise variance $\sigma^2_{Z_{j^*}}\triangleq\min_{j\in\mathcal{J}}\sigma^2_{Z_j}$, where $\sigma^2_{Z_j}\triangleq  10^{-\text{SNR}_E(j)/10}$, $j\in \mathcal{J}$. In other words, the secrecy layer is designed for the best, in terms of signal-to-noise ratio, Eve's channel.  This optimized seed $s$ is then used by the encoder/decoder pair $(e,d)=(e_0 \circ \varphi_s, \psi_s \circ d_0)$, from which we evaluate  $(\mathbf{P}^i_e(e,d),I(M;Z^n_{j}))$, $i\in \mathcal{I}$, $j\in \mathcal{J}$, and  $(\mathbf{P}^{\mathbf{i}}_e(e,d),I(M;Z^n_\mathbf{j}))$, $\mathbf{i}\in \mathcal{I}^n$,~$\mathbf{j}\in \mathcal{J}^n$ in Sections~\ref{comp_s}, \ref{arb_s}.

\subsection{Simulations and examples of code designs for the compound wiretap channel}\label{comp_s}

\subsubsection{Average probability of error analysis}
\begin{figure}[ht]
 \begin{center}
\includegraphics[trim=3.5cm 8cm 2.5cm 9cm,clip,width=0.55\textwidth]{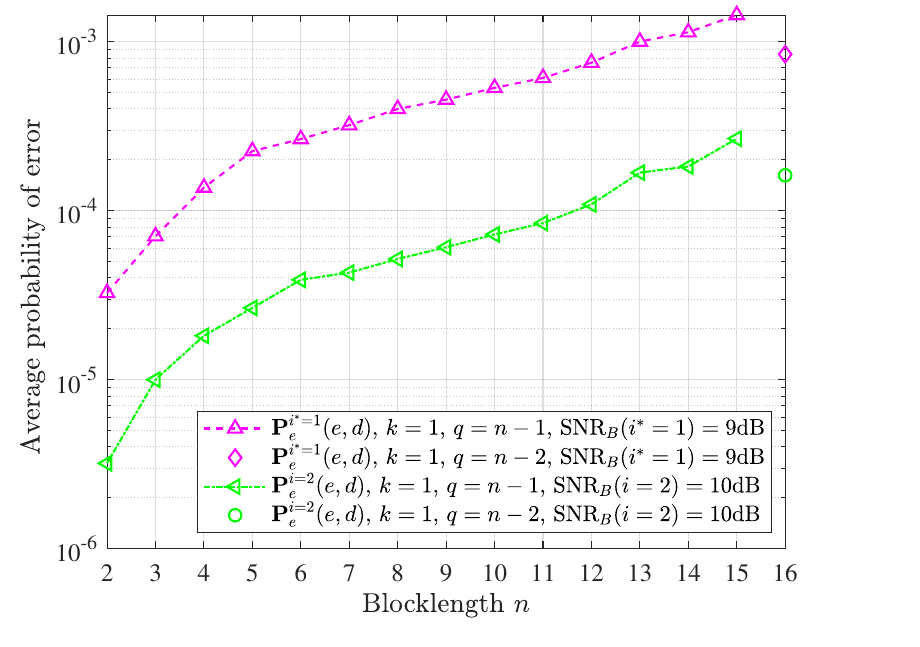} 
\caption{Average probability of error  versus blocklength $n$. }\label{comp_err}
\end{center}
\end{figure}

In our simulations, we consider $\mathcal{I}=\{1,2\}$ and  $\text{SNR}_B(i)\in\{9,10\}\textup{dB}$, $i\in \mathcal{I}$. We evaluate the average probability of error~$\mathbf{P}^{i}_e(e,d)$ for $i \in \mathcal{I}$ as follows. The autoencoder is trained at $\text{SNR}_B(i^*)=9\textup{dB}$, where $i^*=1$.  The message~$M \in \{0,1\}^k$ generated uniformly at random is passed through the trained encoder~$e_0$, which generates the codewords $X^n$ and the channel output~$Y_i^n\triangleq~X^n+N^n_{Y_{i}}$, $i\in\mathcal{I}$, where $N_{Y_{i}}\sim \mathcal{N}(0, \sigma^2_{Y_i})$. Then, the trained decoder $d_0$ forms an estimate $\widehat{M}_i, i\in\mathcal{I}$.  Here, $\sigma^2_{Y_{{i}}}$ is fixed for the entire duration of the transmission. 
We use $5\times10^6$ random messages to evaluate the average probability of error.  Figure~\ref{comp_err} shows the average probability of error $\mathbf{P}^{i}_e(e,d)$  when $k=1$ for $\text{SNR}_B(i^*=1)=9$\textup{dB} and~$\text{SNR}_B(i=2)=10$\textup{dB}.  We observe from Figure~\ref{comp_err} that it is sufficient to design our code for the worst signal-to-noise ratio for Bob, i.e., $\text{SNR}_B(i^*)=9\textup{dB}$. In particular, we observe that, irrespective of what the actual channel is, Bob is able to decode the message with a probability of error smaller than or equal to $\mathbf{P}_e^{i^*=1}(e,d)$. 
\begin{figure}[ht]
\begin{center}
\includegraphics[trim=3.5cm 8cm 2.5cm 9cm,clip,width=0.55\textwidth]{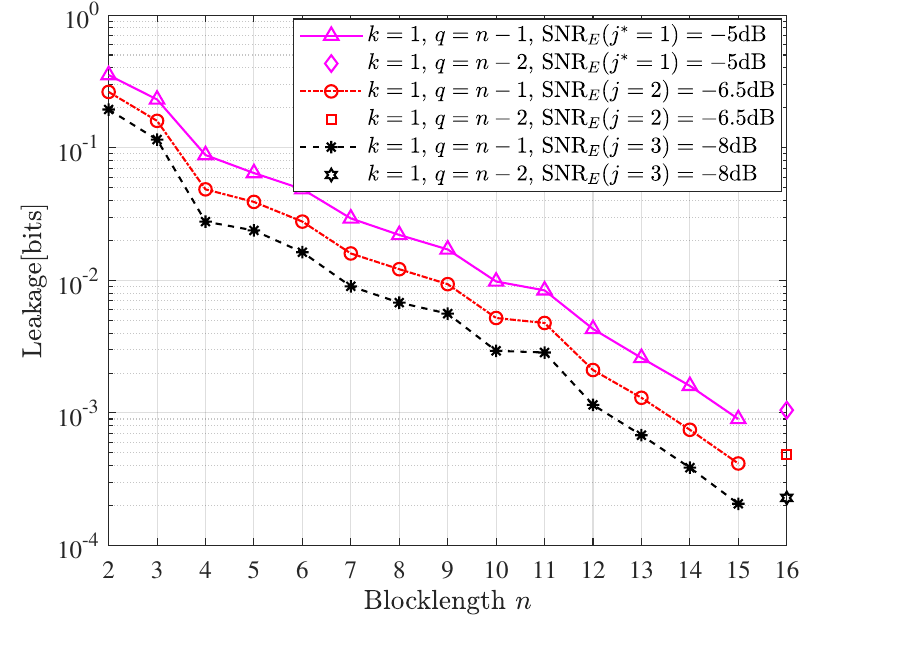} 
\caption{Leakage $\widehat{I}(M;Z_j^n)$ versus blocklength $n$.} \label{comp_lk}
\end{center}
\end{figure}
\subsubsection{Leakage evaluation}
For the simulations, we consider  $\mathcal{J}= \{1,2,3\}$ and $\text{SNR}_E(j)\in~\{-8,-6.5,-5\}\textup{dB}$, $j\in \mathcal{J}$. 
We  compute the leakage $I(M;Z_j^n )$ for $j\in \mathcal{J}$ as in Section~\ref{sim_siso_lk}. The message $M\in \{0,1\}^k$ is passed through the trained encoder $e_0$, which generates the codewords~$X^n$, and the channel output at Eve is $Z_j^n\triangleq X^n+N^n_{Z_j},j\in \mathcal{J}$, where $N_{Z_{j}}\sim \mathcal{N}(0, \sigma^2_{Z_j})$. The noise variance $\sigma^2_{Z_j}$ is fixed for the entire duration of the transmission. 
Figure~\ref{comp_lk} shows the estimated leakage~$\widehat{I}(M;Z_j^n ),$ at $\text{SNR}_E(j=1)=-8\textup{dB}$, $\text{SNR}_E(j=2)=-6.5\textup{dB}$, and~$\text{SNR}_E(j^*=3)=-5\textup{dB}$. From Figure~\ref{comp_lk}, we observe that it is sufficient to design our code for the best signal-to-noise ratio, i.e., $\text{SNR}_E({j^*})=-5\textup{dB}$. In particular, we see that, irrespective of what the actual eavesdropper's channel is, we  always achieve a leakage smaller than or equal to $\widehat{I}(M;Z_{j^*}^n )$, which is also supported  by the following inequality on the leakage. For $j \in \mathcal{J}$ and $j^* \in \argmin_{j\in\mathcal{J}} \sigma^2_{Z_j}$, we have
\begin{align*}
    I(M;Z_j^n) 
    & \leq I(M;Z_j^nZ_{j^*}^n) \\
    & = I(M;Z_{j^*}^n) + I(M;Z_j^n|Z_{j^*}^n) \\
    & = I(M;Z_{j^*}^n),
\end{align*}
where the first inequality holds by the chain rule and nonnegativity of the mutual information, the first equality holds by the chain rule, and the last equality holds because, without loss of generality, one can redefine $Z_j$ such that $Z_j = Z_{j^*} + N$, where $N \sim \mathcal{N}(0, \sigma^2_{Z_j} - \sigma^2_{Z_{j^*}} )$, 
since the distributions $p_{Z_j|X}$ and $p_{Z_{j^*}|X}$ are preserved and the constraints~\eqref{eqc1} and \eqref{eqc2} of the problem  do not depend on the joint distributions $p_{Z_jZ_{j^*}}$.

As an example, from Figures~\ref{comp_err} and~\ref{comp_lk}, we see that for $\text{SNR}_B(i)\in\{9,10\}\textup{dB}$, $i\in \mathcal{I}$, and $\text{SNR}_E(j)\in~\{-8,-6.5,-5\}\textup{dB}$, $j\in \mathcal{J}$, we have designed codes that  show that  the secrecy rate $\frac{1}{8}$ is $(\epsilon= 4.0 \cdot 10^{-4},\delta= 2.2\cdot10^{-2})$-achievable with blocklength $n=8$.

\begin{figure}[ht]
\begin{center}
 \includegraphics[trim=3.5cm 8cm 2.5cm 9cm,clip,width=0.55\textwidth]{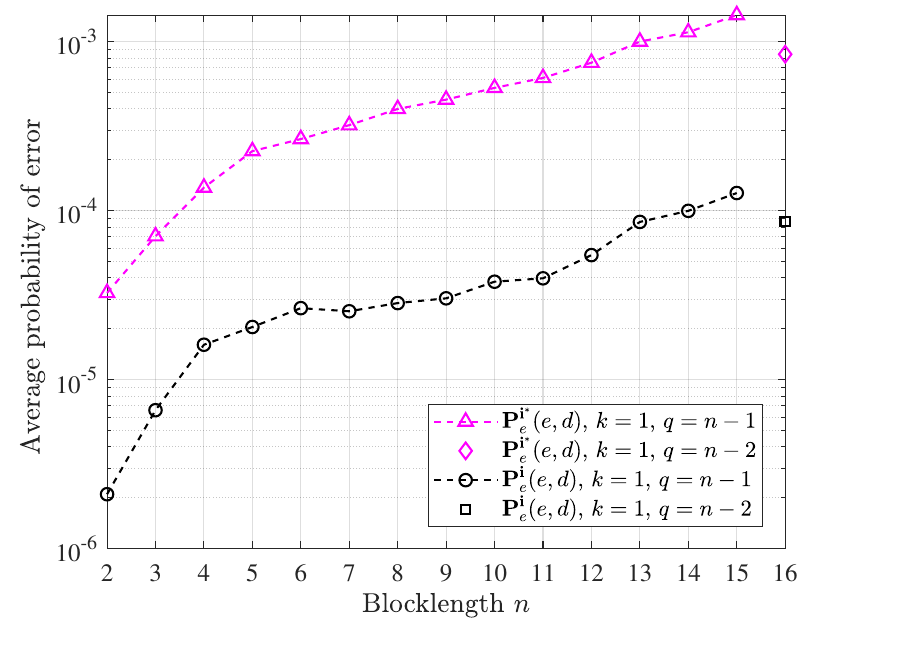} 
\caption{Average probability of error  versus blocklength $n$ when $k=1$ and training $\text{SNR}_B(\mathbf{i}^*)=9$\textup{dB}. } \label{avc_err}
\end{center}
\end{figure}
\subsection{Simulations and examples of code designs for the arbitrarily varying wiretap channel}\label{arb_s}
\subsubsection{Average probability of error analysis}
For the arbitrarily varying channel, we evaluate the probability of error $\mathbf{P}_e^{\mathbf{i}}(e,d)$, $\mathbf{i}\in \mathcal{I}^n$, for $k=1$ in Figure~\ref{avc_err} as follows.  We consider~$\mathcal{I}=~\{1,2,3,\dots,31\}$ and $\text{SNR}_B({i})\in\{9,9.1,9.2,\dots,12\}$, $i \in \mathcal{I}$. The autoencoder is trained at $\text{SNR}_B({i}^*=1)=9\textup{dB}$, where 
the noise variance is fixed for the entire duration of the transmission. The message $M\in\{0,1\}^k$ generated uniformly at random is passed through the trained encoder $e_0$, which generates the codewords~$X^n$ and the channel output at Bob  $Y_{\mathbf{i}}^n\triangleq X^n+N_{Y^n_{\mathbf{i}}}$, $\mathbf{i}\in\mathcal{I}^n$. 
Then, the trained decoder $d_0$ forms an estimate $\widehat{M}_\mathbf{i}$, $\mathbf{i}\in\mathcal{I}^n$. Here, $N_{Y^n_{\mathbf{i}}}$ is a length $n$ vector whose variance of each component is picked uniformly at random from the known uncertainty set~$\{10^{-12\textup{dB}/ 10},10^{-11.9\textup{dB}/ 10},10^{-11.8\textup{dB}/ 10},\dots, 10^{-9\textup{dB} / 10}\}$. For our simulations, the variance of the noise vector $N_{Y^n_\mathbf{i}}$  is fixed for every $50,000$ codewords. The autoencoder is tested  with $5\times 10^6$ random messages for $n>10$ and with $ 10^7$ random messages for $n\leq 10$. Figure~\ref{avc_err} shows that 
even though there is a mismatch between the training signal-to-noise ratio of the encoder/decoder pair and the actual channel, Bob is still able to decode the message with a probability of error smaller than or equal to  $\mathbf{P}_e^{\mathbf{i}^*}(e,d)$, where $\mathbf{i}^*$ is a vector made of $n$ repetitions~of~$i^*$.

\begin{figure}[ht]
\begin{center}
\includegraphics[trim=3.5cm 8cm 2.5cm 9cm,clip,width=0.55\textwidth]{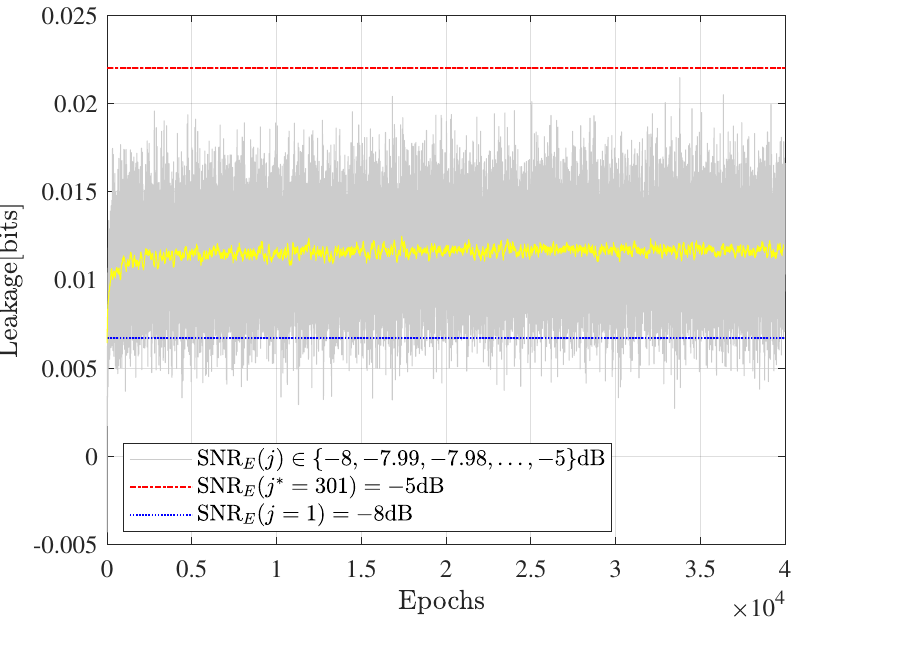} 
\caption{Example of leakage $\widehat{I}(M;Z_{\mathbf{j}}^n)$ versus epochs when $k=1$, $n=8$, $q=n-1$, $s=001101$, and  $j \in \{1,2,3,\dots,301\}$. The grey curve represents the estimated leakage by a  mutual information neural estimator and
the yellow curve represents the $100$-sample moving average of the estimated leakage. The red and blue curves represent the estimated  leakage for $\text{SNR}_E({j^*=301})$ and $\text{SNR}_E({j=1})$, respectively, after convergence.
} \label{avc_lk_1}
\end{center}
\end{figure}
\subsubsection{Leakage evaluation}
For the arbitrarily varying channel, we evaluate the leakage $I(M;Z_{\mathbf{j}}^n )$, for $ \mathbf{j}\in \mathcal{J}^n$, as  in Section~\ref{sim_siso_lk}. The channel output at Eve is~$Z_\mathbf{j}^n\triangleq X^n+N_{Z^n_\mathbf{j}},\mathbf{j}\in \mathcal{J}^n$. In Figure~\ref{avc_lk_1}, we consider $\mathcal{J}= 
\{1,2,3,\dots,301\}$ and~$\text{SNR}_E({j})~\in~\{-8,-7.99,-7.98,\dots,-5\}\textup{dB}$, ${j}\in \mathcal{J}$. Figure~\ref{avc_lk_1} shows the estimated mutual information $\widehat{I}(M;Z_{\mathbf{j}}^n )$ when $k=1$ and $n=8$, where the variance of the noise vector~$N_{Z^n_\mathbf{j}}$ is fixed for $20,000$ codewords per epoch. Here, $N_{Z^n_{\mathbf{j}}}$ is a length $n$ vector whose variance of each component is picked uniformly at random  from the known uncertainty set~$\{10^{5\textup{dB}/ 10},10^{5.01\textup{dB}/ 10},10^{5.02\textup{dB}/ 10},\dots, 10^{8\textup{dB} / 10}\}$.   We can see from Figure~\ref{avc_lk_1} that it is sufficient to design our code for the best signal-to-noise ratio, i.e.,  $\text{SNR}_E({j^*=301})=-5\textup{dB}$. In particular, we observe that, regardless of what the actual channel is, we always achieve a leakage smaller than or equal to $\widehat{I}(M;Z_{\mathbf{j}^*}^n )$, where $\mathbf{j}^*$ is a vector made of $n$ repetitions of $j^*$,  which is also supported by the following inequality on the leakage.

For any $\mathbf{j}\in \mathcal{J}^n$, we have 
$
    I(M;Z_{\mathbf{j}}^n) \leq I(M;Z_{\mathbf{j}^*}^n),
$ because, similar to the compound setting, without loss of generality, one can redefine $Z_{\mathbf{j}}^n$ such that $ M-Z_{\mathbf{j}}^n- Z_{\mathbf{j}^*}^n$ forms a Markov~chain. 
\vspace{-0em}
\begin{figure}[h]
\begin{center}
\includegraphics[trim=3.5cm 8cm 2.5cm 9cm,clip,width=0.55\textwidth]{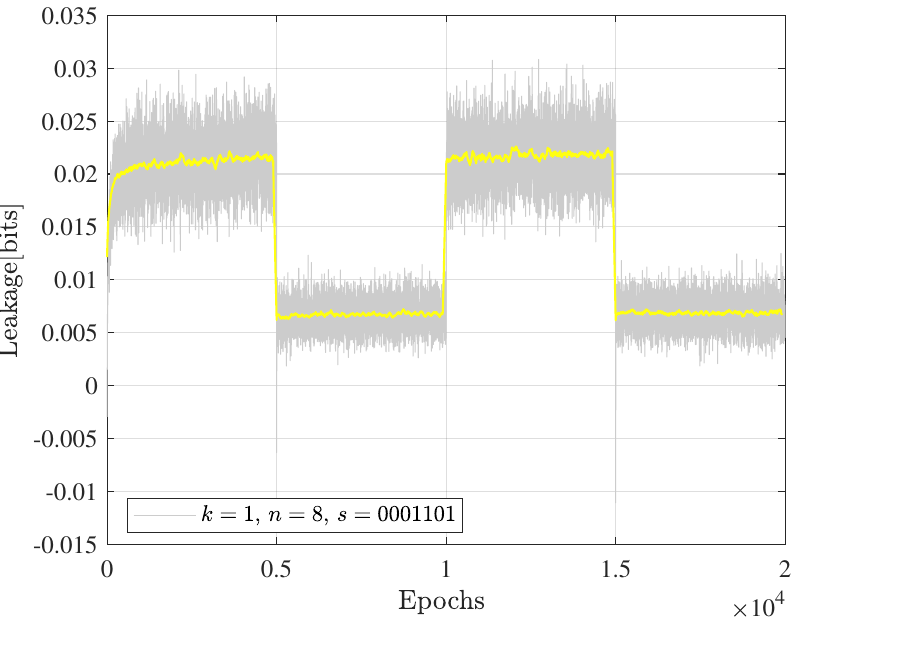} 
\caption{Example of leakage $\widehat{I}(M;Z_{\mathbf{j}}^n)$ versus epochs when $q=n-1$ and $\text{SNR}_E(j)\in\{-8,-5\}$, $j \in \{1,2\}$. 
The yellow curve represents the $100$-sample moving average of the estimated leakage. 
}\label{avc_lk_2}
\end{center}
\end{figure}

Figure~\ref{avc_lk_2} shows the estimated mutual information~$\widehat{I}(M;Z_{\mathbf{j}}^n )$ for $k=1$ and $n=8$, when $\text{SNR}_E\in\{-8,-5\}\textup{dB}$. Here, each component of the noise vector $N_{Z^n_{\mathbf{j}}}$ has fixed variance, which alternatively takes the values $10^{5\textup{dB}/10}$ and $10^{8\textup{dB}/10}$,  for every $5000$ epochs with $20,000$ codewords per epoch. In other words, each component of the noise vector has variance ${10^{5\textup{dB}/10}}$ for the first $5000$ epochs,  then~${10^{8\textup{dB}/10}}$ for the next $5000$ epochs, and so on. Again, we observe that it is sufficient to consider the worst case for the code design, since the leakage is always upper bounded by the leakage obtained for the best eavesdropper's signal-to-noise ratio, i.e., when $\text{SNR}_E=-5\textup{dB}$.

\section{Concluding Remarks}\label{CR}
We designed short blocklength codes for the Gaussian wiretap channel under an information-theoretic secrecy metric. Our approach consisted in decoupling the reliability and secrecy constraints to offer a simple and modular code design, and to precisely control how small the leakage is. Specifically, we handled the reliability constraint via an autoencoder and the secrecy constraint via hash functions.  We evaluated the performance of our code design through simulations in terms of probability of error at the legitimate receiver and leakage at the eavesdropper for blocklengths smaller than or equal to $128$. Our results provide examples of code designs that outperform the best known achievable secrecy rates obtained non-constructively for the Gaussian wiretap channel. We highlight that our code design method can be applied to any channels  since it does not require the knowledge of the channel model but only the knowledge of input and output channel samples. We also showed that our code design is applicable to settings where uncertainty holds on the channel statistics, e.g.,  compound wiretap channels, and arbitrarily varying wiretap channels.

\appendix
\section{Achievability and converse bounds for the Gaussian wiretap channel}\label{appen_ac_co}

\subsection{Achievability bound for the Gaussian wiretap channel}
The maximal secrecy rate $R(n,\epsilon, \delta)$ achievable by an $\epsilon$-reliable and $\delta$-secure $(n, k,P)$ code is lower bounded as \cite[Theorem 7 and Section IV.C-1]{yang}
\begin{align*}
   R(n,\epsilon, \delta)\geq\frac{1}{n}\log_2\frac{M(\epsilon,n)}{L(n, \delta)},
\end{align*}
with $M(\epsilon,n)$ the number of codewords for a  probability of error~$\epsilon$ and blocklength~$n$ inferred by Shannon's channel coding achievability bound  \cite[Section III.J-4]{polyanskiy}, and $L(n, \delta)$ such~that
\begin{align}
    \sqrt{L(n, \delta)}\triangleq \min_{\gamma}\frac{\sqrt{\gamma \mathbb{E}[\exp (-\vert B_n-\log \gamma \vert)]}}{2(\delta+\mathbb{E}[\exp (-\vert B_n-\log \gamma \vert^+)]-1)},
\end{align}
where the minimization is over all $\gamma>0$ such that the denominator is positive, and 
\begin{align*}
    B_n\triangleq \frac{n}{2}\log_2\Big(1+\frac{P}{\sigma^2_Z}\Big)+\frac{\log_2 e}{2}\sum_{t=1}^{n}\Big(1-\frac{(\sqrt{P}Z_t-\sqrt{\sigma^2_Z})^2}{P+\sigma^2_Z}\Big),
\end{align*}
where $Z_t$, $t\in \{ 1, \ldots, n\}$, are i.i.d. according to the standard normal distribution.

\subsection{Converse bound for the Gaussian wiretap channel}
An  $\epsilon$-reliable and $\delta$-secure $(n, k, P)$ code for the wiretap channel $(\mathcal{X},P_{YZ\vert X}, \mathcal{Y}\times\mathcal{Z})$ satisfies \cite[Theorem 12 and Section IV.C-3]{yang}
\begin{align}
    2^k \leq \inf_{\tau\in(0,1-\epsilon-\delta)}\frac{\tau+\delta}{\tau  \beta_{1-\epsilon-\delta-\tau}(P_{X^nY^nZ^n},P_{X^nZ^n}Q_{Y^n\vert Z^n})},
\end{align}
where $P_{X^nY^nZ^n}$ denotes the joint probability distribution induced by the code and for $Q_{Y^n\vert Z^n}$ as in \cite[Eq. (129)]{yang} $$\beta_{1-\epsilon-\delta-\tau}(P_{X^nY^nZ^n},P_{X^nZ^n}Q_{Y^n\vert Z^n}) \geq \mathbb{P}[\bar{D}_{n+1}\geq \bar{\gamma}],$$ 
where 
\begin{align*}
    \bar{D}_{n+1}&\triangleq (n+1)C_s\\
    &+\frac{\log_2 e}{2}\sum^{n+1}_{t=1}\Bigg(\frac{{N}_{Z_t}^2}{\sigma^2_Z}-\frac{(\bar{N}_{Z_t}-c_0({N}_{Z_t}+\sqrt{P}))^2}{P+\sigma^2_Z}\\&+\frac{\bar{N}_{Z_t}^2}{P+\sigma^2_Y}-\frac{(c_1{N}_{Z_t}+c_0\bar{N}_{Z_t}-c_0^2\sqrt{P})^2}{\sigma^2_Y}\Bigg),
\end{align*}
with ${N}_{Z_t}\sim \mathcal{N}(0,\sigma^2_Z)$, $\bar{N}_{Z_t}\sim \mathcal{N}(0,P+\sigma^2_Y)$, $C_s \triangleq \frac{1}{2} \log_2 \frac{1+P /\sigma_Y^2}{1+P / \sigma_Z^2}$, and $c_0\triangleq \sqrt{\frac{\sigma^2_Z-\sigma^2_Y}{P+\sigma^2_Z}}$, $c_1\triangleq \frac{P+\sigma^2_Y}{P+\sigma^2_Z}$, and the threshold $\bar{\gamma}$ satisfies $\mathbb{P}[\bar{B}_{n+1}\geq \bar{\gamma}]=1-\epsilon-\delta-\tau$ with
\begin{align*}
    \bar{B}_{n+1}&\triangleq (n+1)C_s+\frac{\log_2 e}{2}\sum_{t=1}^{n+1}\Bigg(\frac{(N_{Y_t}+\bar{N}_{Y_t})^2}{\sigma^2_Z}-\frac{N_{Y_t}^2}{\sigma^2_Y}\\&+\frac{(\sqrt{P}+N_{Y_t})^2}{P+\sigma^2_Y}-\frac{(\sqrt{P}+N_{Y_t}+\bar{N}_{Y_t})^2}{P+\sigma^2_Z}\Bigg),
\end{align*}
where $N_{Y_t}\sim \mathcal{N}(0,\sigma^2_Y)$ and $\bar{N}_{Y_t}\sim \mathcal{N}(0,\sigma^2_Z-\sigma^2_Y)$, $t\in \{ 1, \ldots, n+1\}$, are independent and identically distributed.
 
\bibliographystyle{IEEEtran}
\bibliography{poly}

\end{document}